\documentclass[prb,amsfonts,amssymb,floats,twocolumn,aps]{revtex4}

\usepackage[final,dvips]{epsfig}
\usepackage{amsmath}
\usepackage{float}
\usepackage[caption = false]{subfig}
\usepackage{feynmp}
\usepackage{graphicx}
\usepackage{color}
\usepackage{pstricks,pst-grad,pstricks-add}
\usepackage{pst-plot,pst-math}

\newcommand{\w}{\omega}
\newcommand{\qq}{q}     
\newcommand{\qqc}{q_c}  
\newcommand{\gk}{\gamma_{\vec{k}}}
\newcommand{\lm}{\lambda}
\newcommand{\Ja}{J_3}   
\newcommand{\Jb}{J_4}
\newcommand{\hz}{h^z}
\newcommand{\A}{J_1}	
\newcommand{\Ap}{J_2}
\newcommand{\Lm}{\Lambda} 

\newcommand{\ii}{\imath}   

\newcommand{\EE}{E_{0}}  
\newcommand{\EEz}{E_{00}}  
\newcommand{\EEo}{E_{01}}  

\newcommand{\MM}{M_{\rm st}}  
\newcommand{\MMz}{M_{{\rm st}0}}  
\newcommand{\MMo}{M_{{\rm st}1}}  
\newcommand{\MMt}{M_{{\rm st}2}}  

\newcommand{\tb}{\tilde{\tau}}
\newcommand{\ut}{u}
\newcommand{\vt}{v}
\newcommand{\kk}{\vec{k}}
\newcommand{\wt}{\tilde{\omega}}
\newcommand{\wtk}{\tilde{\omega}_{\vec{k}}}
\newcommand{\Wt}{\tilde{\Omega}}
\newcommand{\Wtk}{\tilde{\Omega}_{\vec{k}}}

\newcommand{\vcxa}{\Phi_{31}^{a}}
\newcommand{\vcxb}{\Phi_{32}^{a}}
\newcommand{\vcxc}{\Phi_{33}^{a}}
\newcommand{\vcza}{\Phi_{31}^{z}}
\newcommand{\vczb}{\Phi_{32}^{z}}

\newcommand{\vqaza}{\Phi_{41}^{az}}
\newcommand{\vqazb}{\Phi_{42}^{az}}
\newcommand{\vqazc}{\Phi_{43}^{az}}
\newcommand{\vqazd}{\Phi_{44}^{az}}
\newcommand{\vqaze}{\Phi_{45}^{az}}
\newcommand{\vqazf}{\Phi_{46}^{az}}

\newcommand{\vqza}{\Phi_{41}^{z}}
\newcommand{\vqzb}{\Phi_{42}^{z}}
\newcommand{\vqzc}{\Phi_{43}^{z}}

\newcommand{\vqaba}{\Phi_{41}^{ab}}
\newcommand{\vqabb}{\Phi_{42}^{ab}}
\newcommand{\vqabc}{\Phi_{43}^{ab}}
\newcommand{\vqabd}{\Phi_{44}^{ab}}

\begin{document}

\title[]
{
Non-linear bond-operator theory and $1/d$ expansion for coupled-dimer magnets II: \\
Antiferromagnetic phase and quantum phase transition
}
\author{Darshan G. Joshi}
\author{Matthias Vojta}
\affiliation{Institut f\"ur Theoretische Physik,
Technische Universit\"at Dresden, 01062 Dresden, Germany}

\date{\today}

\begin{abstract}
We extend to magnetically ordered phases a recently developed expansion in $1/d$ for coupled-dimer Heisenberg magnets, where $d$ is the number of space dimensions. This extension utilizes generalized bond operators describing spin excitations on top of a reference state involving triplet condensates.
We explicitly consider a model of dimers on a hypercubic lattice which displays, in addition to the paramagnetic singlet phase, a collinear antiferromagnetic phase for which we calculate static and dynamic observables at zero temperature.
In particular, we show that the $1/d$ expansion smoothly connects the paramagnetic and antiferromagnetic phases and produces sensible results at and near the quantum phase transition point. Among others, we determine the dispersion and spectral-weight distribution of the amplitude (i.e. Higgs) mode of the ordered phase.
In the limit of vanishing intra-dimer coupling, we connect our approach to spin-wave theory.
\end{abstract}

\pacs{}

\maketitle


\section{Introduction}
\label{sec:intro}

Systematic expansions for many-body systems play an important role in theoretical physics, because (i) they allow one to make controlled statements in certain well-defined limits in parameter space, and (ii) they may be extrapolated to cover a large part of parameter space if sufficiently high orders are used. However, identifying a suitable expansion parameter in systems with strong interactions, such as spin systems, is a non-trivial problem. Frequently used expansions utilize artificial small parameters such as $1/N$, where $N$ is the number of flavors or order-parameter components, $1/S$, where $S$ is the spin size, or $\epsilon=d-d_c$, the deviation of the number of space dimensions $d$ from a critical dimension $d_c$.\cite{auerbach,zinnjus,oitmaa}

In a recent paper,\cite{i} henceforth referred to as I, we have developed a $1/d$ expansion for an important class of spin models, namely coupled-dimer Heisenberg magnets.\cite{ssbook,ss_natph,gia08} Such magnets consist of strongly coupled pairs (dimers) of quantum spins which themselves are connected by weaker inter-dimer couplings. For spins 1/2 the general Hamiltonian reads
\begin{equation}
\label{h}
\mathcal{H} =
\sum_i J_i \vec{S}_{i1} \cdot \vec{S}_{i2} +
\sum_{ii'mm'} K_{ii'}^{mm'} \vec{S}_{im} \cdot \vec{S}_{i'm'}
\end{equation}
where the indices $i,i'$ refer to sites on a regular lattice of dimers, and $m,m'=1,2$
refer to the individual spins on each dimer.
In dimensions $d\geq 2$ and for antiferromagnetic (AF) interactions, coupled-dimer models typically display a quantum phase transition (QPT) between a paramagnetic ground state, realized at small $K/J$, and an AF ground state, realized at large $K/J$.

The expansion developed in I is based on a bond-operator representation of the dimer Hilbert space. In contrast to the original mean-field-based linear bond-operator theory,\cite{bondop} it employs an exact projection scheme which, in large space dimensions, enables a perturbative treatment of the non-linear Hamiltonian terms. In I, we have employed this expansion to calculate observables in the paramagnetic phase of a hypercubic-lattice dimer model in a systematic expansion in $1/d$ up to the transition point.

In this paper, we extend the $1/d$ expansion to magnetically ordered phases. Starting from a suitable large-$d$ reference state, i.e., a dimer product state which involves a triplet condensate, we derive a generalized bond-operator Hamiltonian describing fluctuations on top of this reference state.\cite{sommer,penc}
This Hamiltonian is then used to generate a $1/d$ expansion for the reference state itself as well as for thermodynamic and spectral properties. Connections between our non-linear bond-operator theory and non-linear spin-wave theory as known from the literature will be highlighted.

As in I, we demonstrate the approach for a hypercubic-lattice coupled-dimer model whose ordered state is a collinear antiferromagnet. We determine the order parameter as well as dispersion and spectral weight of both transverse (i.e. Goldstone) and longitudinal (i.e. Higgs) excitations in this phase. We show that the present $1/d$ expansion smoothly connects to the paramagnetic phase of the model, with a continuous QPT between the two.
Our theory thus succeeds in consistently describing, beyond the level of Gaussian fluctuations, both Goldstone and critical modes in a system with a condensate order parameter -- this is highly non-trivial considering that standard approaches to the interacting-boson problem, like the Hartree-Fock and Popov approximations,\cite{hfp1,hfp2} as well as previous approaches to coupled-dimer magnets\cite{chub95,kotov,kotov_ordered} fail in this respect.
Altogether, this turns the popular bond-operator formalism into a controlled and systematic theory.

\subsection{$1/d$ expansion, Goldstone modes, and quantum phase transitions}
\label{sec:mfchar}

The utility of the small parameter $1/d$ guarantees sensible and consistent results across the entire phase diagram -- this distinguishes our approach from earlier refinements of bond-operator theory\cite{kotov} or alternative microscopic approaches to the Heisenberg bilayer model.\cite{chub95} In particular, the minimum energy of transverse spin fluctuations in the ordered phase of an SU(2) symmetric coupled-dimer model, being zero at any $d$ due to Goldstone's theorem, is zero to all orders in the $1/d$ expansion.\cite{otsuki}

As discussed in I, the $1/d$ expansion can also access the vicinity of the quantum critical point despite the presence of critical singularities:
First, critical exponents necessarily take mean-field values to all orders in the $1/d$ expansion.
Second, observables which are analytic at criticality are amenable to a $1/d$ expansion even across the QPT. In I, this was demonstrated for the excitation gap $\Delta$ of the disordered state which varies with the distance $t$ to criticality as $\Delta \propto t^{\nu z}$ with $\nu=1/2$, $z=1$, hence $\Delta^2 \propto t$ is analytic. Here we shall determine, e.g., the order parameter $\MM$ which follows $\MM \propto (-t)^\beta$ with $\beta=1/2$, hence $\MM^2 \propto (-t)$ is analytic as well. In the above, $\nu$, $z$, and $\beta$ are the correlation-length, dynamic, and order-parameter exponents, respectively.

We note that applying the $1/d$ expansion near the QPT can also be used to extract the coefficients of a $\phi^4$ order-parameter field theory, which then may be employed to analyze critical properties. As we are able to {\em directly} calculate observables at and near criticality, we shall not discuss this route further.

\subsection{Model and summary of results}
\label{sec:introsumm}

We summarize our main results obtained for the coupled-dimer model on a $d$-dimensional hypercubic lattice, with
\begin{align}
\label{hh}
\mathcal{H} &=
J \sum_i \vec{S}_{i1} \cdot \vec{S}_{i2} +
\sum_{\langle ii'\rangle} (K^{11} \vec{S}_{i1} \cdot \vec{S}_{i'1} + K^{22} \vec{S}_{i2} \cdot \vec{S}_{i'2})
\nonumber \\
&+ \hz \sum_{i} e^{i\vec{Q} \cdot R_{i}} (S_{i1}^{z} - S_{i2}^{z}) \,.
\end{align}
Here, $\sum_{\langle ii'\rangle}$ denotes a summation over pairs of nearest-neighbor dimer
sites, and we will exclusively consider the symmetric case with $K^{11}=K^{22}\equiv K$.
We have added a staggered field $\hz$ which couples to the collinear AF order parameter at $\vec{Q} = (\pi,\pi,\ldots)$.

As in I, the ratio between inter-dimer and intra-dimer coupling is parameterized by
\begin{equation}
\qq = \frac{Kd}{J}
\end{equation}
which ensures a non-trivial competition between these interactions in the limit $d\to\infty$ at fixed $\qq$. In $d=2$, where Eq.~\eqref{hh} represents the much-studied bilayer Heisenberg model, the transition between the paramagnetic and collinear AF phases occurs at\cite{sandvik06} $\qqc = 0.793$.

From our large-$d$ expansion in the AF phase, we find the QPT to be located at $\hz=0$ and
\begin{equation}
\label{qqcexp}
\qqc = \frac{1}{2} + \frac{3}{16} \frac{1}{d} + \mathcal{O}\Big(\frac{1}{d^2}\Big)\,,
\end{equation}
identical to the corresponding result obtained in I for the paramagnetic phase.
At $\hz=0$ the staggered magnetization per dimer follows
\begin{equation}
\label{mstintro}
\MM^{2} =
\frac{4\qq^{2}-1}{4\qq^{2}} - \frac{1}{d}\left[ \frac{5 (2\qq+1)^{2}}{256 \qq^{6}} +1 \right] \frac{2 \qq^{2}}{(2\qq+1)^{2}} +\mathcal{O}\left(\frac{1}{d^2}\right)
\end{equation}
and vanishes at the critical point as
\begin{equation}
\label{delzintro}
\MM = \left[ 2 + \frac{5}{3d} +\mathcal{O}\left(\frac{1}{d^2}\right) \right] \sqrt{\qq-\qqc}\,.
\end{equation}
The gap $\Delta_z$ of the longitudinal (Higgs) mode is given by
\begin{align}
\notag
\frac{\Delta_{z}^{2}}{J^2} = 4\qq^{2} -1 &+ \frac{1}{32d} \Big[ -\frac{1}{\qq^2} -\frac{16}{(2\qq+1)^2} + \frac{48}{2\qq+1} \\
&+ \frac{192}{12\qq^2 +1} -96 +16\qq \Big]
+ \mathcal{O}\left(\frac{1}{d^2}\right).
\end{align}
It closes at the critical point as
\begin{equation}
\label{delzintro}
\frac{\Delta_{z}}{J} = \left[ 2 - \frac{5}{8 d} +\mathcal{O}\left(\frac{1}{d^2}\right)\right] \sqrt{\qq-\qqc}\,.
\end{equation}
Both longitudinal and transverse modes have the same velocity at criticality, with the $1/d$ expansion
\begin{equation}
\label{cintro}
\frac{c}{J} = \frac{1}{\sqrt{2}} + \frac{5}{16\sqrt{2} d} + \mathcal{O}\Big(\frac{1}{d^2}\Big).
\end{equation}

\subsection{Outline}

The body of the paper is organized as follows:
Section~\ref{sec:bond} describes the generalization of the bond-operator approach to magnetically ordered phases. In Section~\ref{sec:ham} we apply this formalism to the collinear phase of the hypercubic bilayer model, where we define a suitable reference state, derive an exact interacting bond-operator Hamiltonian for its excitations, and discuss the strategy to construct a $1/d$ expansion. The explicit calculation of observables, order by order in $1/d$, is demonstrated in Section~\ref{sec:expa}. Section~\ref{sec:spinw} finally highlights the similarities and differences between non-linear spin-wave theory and our approach when applied deep in the ordered phase.

A concluding section closes the paper, and various appendices are devoted to technical details.


\section{Bond operators for ordered phases}
\label{sec:bond}

Sachdev and Bhatt \cite{bondop} devised bond-operator mean-field theory as an efficient slave particle-description for the quantum paramagnetic phase of coupled-dimer magnets \eqref{h}. In the original formulation, the singlet state on each dimer is ``condensed'', and triplet excitations (later dubbed ``triplons'') on top of this singlet state are treated as non-interacting bosons. Bond-operator theory has been generalized to magnetically ordered phases by using triplet condensates.\cite{sommer,penc,MatsumotoNormandPRL} Here we formulate this generalization such that it can be combined with an exact projection scheme suitable for the $1/d$ expansion.

We denote the four basis states on each dimer $i$ by $|t_k\rangle_i$,
$k=0, \ldots, 3$, where
$|t_0\rangle=
(|\uparrow\downarrow\rangle -|\downarrow\uparrow\rangle)/\sqrt{2}$ is the spin-0 singlet
state, and
$|t_1\rangle =
(-|\uparrow\uparrow\rangle+|\downarrow\downarrow\rangle)/\sqrt{2}$,
$|t_2\rangle =
\ii(|\uparrow\uparrow\rangle+|\downarrow\downarrow\rangle)/\sqrt{2}$,
$|t_3\rangle =
(|\uparrow\downarrow\rangle+|\downarrow\uparrow\rangle)/\sqrt{2}$
correspond to the spin-1 triplet, and $\ii$ is the imaginary unit.

\subsection{General Hilbert-space rotation}

While the paramagnetic phase of a coupled-dimer model can be conveniently accessed from a state involving a product of singlets, $|\psi_0\rangle = \prod_i |t_0\rangle_i$, magnetically ordered phases require a reference state with broken SU(2) spin symmetry.
For a consistent description of excitations within a modified bond-operator formalism it is convenient to perform an SU(4) basis rotation in the Hilbert space of each dimer.\cite{sommer} The most general form reads
\begin{equation}
\label{trafo}
|\tilde{t}_k\rangle_i = U_{kk'}^{(i)} |t_{k'}\rangle_i,~
~~~~(k,k'=0,\ldots,3).
\end{equation}
The rotation should be chosen such that $|\tilde{\psi}_0\rangle = \prod_i |\tilde{t}_0\rangle_i$ is a suitable reference state which replaces the singlet product state $|\psi_0\rangle$. For instance, a local N\'eel state polarized along $z$ is obtained from $|\tilde{t}_0\rangle = (|t_0\rangle + |t_3\rangle)/\sqrt{2} = |\uparrow\downarrow\rangle$.

Spin operators $\vec{S}_{im}$ can be represented in terms of transitions between the states $|t_k\rangle_i$ of a dimer,
\begin{equation}
\label{strans}
{S}_{im}^\alpha = \sum_{kk'} s_{kk'}^{\alpha m} |t_k\rangle_i {\phantom .}_i\langle t_{k'}|,
\end{equation}
with $4\times4$ matrices $s^{\alpha m}$ for the spin components $S^\alpha$ ($\alpha =
x,y,z\equiv 1,2,3$) of the $m=1,2$ spins:
\begin{eqnarray}
s^{x1,2} &=& \frac{1}{2} \left(
\begin{array}{cccc}
0 & \pm 1 & 0 & 0 \\
\pm 1 & 0 & 0 & 0 \\
0 & 0 & 0 & -\ii \\
0 & 0 & \ii & 0
\end{array}
\right), \nonumber\\
s^{y1,2} &=& \frac{1}{2} \left(
\begin{array}{cccc}
0 & 0 & \pm 1 & 0 \\
0 & 0 & 0 & \ii \\
\pm 1 & 0 & 0 & 0 \\
0 & -\ii & 0 & 0
\end{array}
\right), \nonumber\\
s^{z1,2} &=& \frac{1}{2} \left(
\begin{array}{cccc}
0 & 0 & 0 & \pm 1 \\
0 & 0 & -\ii & 0 \\
0 & \ii & 0 & 0 \\
\pm 1 & 0 & 0 & 0
\end{array}
\right).
\label{smat}
\end{eqnarray}
This is of course equivalent to the bond-operator representation of Sachdev and Bhatt,\cite{bondop} written in terms of transition operators:
\begin{equation}
S_{i1,2}^\alpha = \frac{1}{2}\left(\pm |t_0\rangle_i {\phantom .}_i\langle t_{\alpha}|
              \pm |t_{\alpha}\rangle_i {\phantom .}_i\langle t_{0}|
          - \ii\epsilon_{\alpha\beta\gamma} |t_{\beta}\rangle_i {\phantom .}_i\langle t_{\gamma}| \right).
\nonumber
\end{equation}
After the basis rotation \eqref{trafo}, Eq.~\eqref{strans} becomes
\begin{equation}
\label{strans2}
{S}_{im}^\alpha = \sum_{kk'} \tilde{s}_{i,kk'}^{\alpha m} |\tilde{t}_k\rangle_i {\phantom .}_i\langle \tilde{t}_{k'}|,
\end{equation}
with the transformed spin matrices now being in general site-dependent:
\begin{equation}
\label{smat2}
\tilde{s}_{i,kk'}^{\alpha m} = \sum_{ll'} (U^\dagger)_{lk}^{(i)} s_{ll'}^{\alpha m} U_{k'l'}^{(i)}.
\end{equation}

\subsection{Excitations and projection}

The next step is to introduce bosonic operators $\tilde{t}_{i\alpha}$ ($\alpha=1,2,3$) for local excitations w.r.t. the reference state $|\tilde{t}_0\rangle_i$,
\begin{equation}
\label{gbond}
|\tilde{t}_\alpha\rangle_i = \tilde{t}_{i\alpha}^\dagger |\tilde{t}_0\rangle_i.
\end{equation}
In the untransformed case, the $\tilde{t}_{i\alpha}^\dagger$ are the triplon bond operators as used in Refs.~\onlinecite{kotov,i}, and we will continue to refer to them as (generalized) triplons.
These operators obey a hard-core constraint,
\begin{equation}
\label{hardcore}
\sum_{\alpha=1}^3  \tilde{t}^\dagger_{i\alpha} \tilde{t}_{i\alpha} \leq 1.
\end{equation}
As discussed in some detail in I, this constraint is efficiently implemented using projection operators $P_i$ which suppress all matrix elements of observables between the physical and unphysical parts of the Hilbert space, i.e., prevent the creation of more than one triplon excitation per dimer site $i$.
As in Refs.~\onlinecite{i,alter,alter2} we choose projectors
\begin{equation}
\label{proj}
P_i = 1- \sum_\gamma \tilde{t}_{i\gamma}^\dagger \tilde{t}_{i\gamma} \,.
\end{equation}
With the help of the $P_i$ the transitions between the dimer states can now be written in terms of the generalized bond operators \eqref{gbond} as follows:
\begin{align}
|\tilde{t}_0\rangle_i {\phantom .}_i\langle \tilde{t}_0|          &= P_i,\notag\\
|\tilde{t}_\alpha\rangle_i {\phantom .}_i\langle \tilde{t}_0|     &= \tilde{t}_{i\alpha}^\dagger P_i,\notag\\
|\tilde{t}_0\rangle_i {\phantom .}_i\langle \tilde{t}_\alpha|     &= P_i \tilde{t}_{i\alpha},\notag\\
|\tilde{t}_\alpha\rangle_i {\phantom .}_i\langle \tilde{t}_\beta| &= \tilde{t}_{i\alpha}^\dagger \tilde{t}_{i\beta}.
\label{transbond}
\end{align}
Inserted in \eqref{smat2}, these relations allow to re-write the Hamiltonian and other observables in terms of the $\tilde{t}_{i\alpha}$ bosons. In particular, the spin operators, when expressed via the $\tilde{t}_{i\alpha}$, obey standard spin commutation within the physical Hilbert space defined by Eq.~\eqref{hardcore}.


\section{Reference state and Hamiltonian}
\label{sec:ham}

In this section we turn to the hypercubic-lattice coupled-dimer model \eqref{hh} and describe how to set-up the $1/d$ expansion for the AF ordered phase.
This requires (i) to define a suitable reference state and a corresponding Hilbert-space rotation, (ii) to express the Hamiltonian in the generalized bond operators, (iii) to perform a Bogoliubov transformation for the leading-order bilinear part, and (iv) to express and normal-order the remaining Hamiltonian in terms of the Bogoliubov-transformed triplon operators.
These steps, together with a discussion of the expansion strategy, can be found in the following subsections.

\subsection{Reference product state}
\label{sec:refstate}

For dominant AF inter-dimer interaction $K$, the hypercubic-lattice model \eqref{hh} realizes a collinear N\'{e}el state on each of the $m=1,2$ ``layers'', with the two layers having opposite spin orientation.
Assuming that the staggered magnetization of the ordered state points along $\hat{z}$, its description requires an alternating linear combination of singlet and $z$-triplet, i.e., we choose a Hilbert-space rotation involving a single real condensate parameter $\lm$:
\begin{align}
\label{rotate}
|\tilde{t}_0\rangle_i &= (|t_0\rangle_i + \lm_i |t_3\rangle_i) / \sqrt{1+\lm^2} \,,\\
\label{rotate2}
|\tilde{t}_3\rangle_i &= (|t_3\rangle_i - \lm_i |t_0\rangle_i) / \sqrt{1+\lm^2} \,,\\
\label{rotate3}
|\tilde{t}_1\rangle_i &= |t_1\rangle_i,~|\tilde{t}_2\rangle_i = |t_2\rangle_i\,,
\end{align}
with $\lm_i = \lm e^{i{\vec Q}\cdot{\vec r}_i} = \pm \lm$, or equivalently
\begin{equation}
U^{(i)} = \left(
            \begin{array}{cccc}
              c\lm_i & 0 & 0 & s\lm_i \\
              0 & 1 & 0 & 0 \\
              0 & 0 & 1 & 0 \\
              -s\lm_i & 0 & 0 & c\lm_i \\
            \end{array}
          \right),
\end{equation}
with $s\lm_i = \sin \tan^{-1}\lm_i$ and $c\lambda_i = \cos \tan^{-1} \lm_i$.
Apparently, $|\tilde{t}_0\rangle_i$ smoothly interpolates between a singlet for $\lm=0$ and a $\hat{z}$-oriented N\'{e}el configuration for $\lm=\pm 1$.
In the latter case, the excitations created by the $\tilde{t}^\dagger_{i\alpha}$ operators are easily interpreted: $\tilde{t}^\dagger_{i1,2} \equiv \tilde{t}^\dagger_{ix,y}$ correspond to {\em transverse} (or single spin-flip) excitations which will yield the Goldstone modes of the ordered phase. In contrast, $\tilde{t}^\dagger_{i3} \equiv \tilde{t}^\dagger_{iz}$ is a {\em longitudinal} excitation: for $\lm=1$ we have $|\tilde{t}_0\rangle = |\uparrow\downarrow\rangle$ and $|\tilde{t}_3\rangle = |\downarrow\uparrow\rangle$, i.e., $\tilde{t}^\dagger_{i3}$ flips both dimer spins. The interpretation of the modes will substantiated by the dispersion results obtained below.

The value of the rotation (or condensate) parameter $\lm$ is left unspecified at this point; it depends on model parameters and will acquire a $1/d$ expansion, to be described below. This is similar to the behavior of the reference state in spin-wave theory for non-collinear states, e.g., for an antiferromagnet in a uniform field: Here the moment orientation receives corrections at every order in $1/S$.

We note that the reference state $|\tilde{\psi}_0\rangle = \prod_i |\tilde{t}_0\rangle_i$ is suitable for an applied staggered field along $\hat{z}$, but cannot describe the physics in a uniform field, as it yields zero net magnetization. Linear bond-operator theory in the presence of a uniform field using canted states has been described in Ref.~\onlinecite{sommer}; we leave the corresponding $1/d$ expansion for future work.

\subsection{Real-space bond-operator Hamiltonian}

The Hamiltonian of the model \eqref{hh} can be expressed using the rotated bond
operators $\tilde{t}_{i\alpha}$, with arbitrary condensate parameter $\lambda$. Inserting the projectors $P_i$ \eqref{proj}, the resulting Hamiltonian can be split as follows:
\begin{align}
\label{hhexp}
\mathcal{H} = \mathcal{H}_0 + \mathcal{H}_1 + \mathcal{H}_2 + \mathcal{H}_3 + \mathcal{H}_4 + \mathcal{H}_5 + \mathcal{H}_6
\end{align}
where the $\mathcal{H}_n(\lm)$ contain $n$ triplon operators $\tilde{t}_{i\alpha}$ and explicitly depend on the reference-state parameter $\lm$. In contrast to the calculation in the paramagnetic phase, here all $\mathcal{H}_n$ with odd $n$ are non-zero even for a symmetric system with $K^{11}=K^{22}$.

We list the terms up to order four, as these are required for the following $1/d$ expansion (recall $\lm_i = \lm e^{i{\vec Q}\cdot{\vec r}_i}$):

\begin{widetext}
\begin{align}
\label{H0}
\mathcal{H}_0 &= -\frac{N J (3-\lambda^2)}{4(1+\lambda^2)} -
\frac{2NKd\lambda^2}{(1+\lambda^2)^2} +
\frac{2 N \hz \lambda}{1+\lambda^2}\,,  \\
\label{H1}
\mathcal{H}_1 &= \sum_{i} e^{i \vec{Q}\cdot\vec{r_i}} \left[
\frac{\lambda J}{1+\lambda^2} - \frac{2Kd \lambda (1-\lambda^2)}{(1+\lambda^2)^2} +
\frac{\hz (1-\lambda^2)}{1+\lambda^2} \right]
(\tilde{t}_{iz}^{\dagger} + \tilde{t}_{iz})\,, \\
\label{H2}
\mathcal{H}_2 &= \sum_{i,a} \left[ \frac{J}{1+\lambda^2} - \frac{2\lambda \hz}{1+\lambda^2} +
\frac{4 Kd \lambda^2}{(1+\lambda^2)^2} \right] \tilde{t}_{ia}^{\dagger} \tilde{t}_{ia} +
\sum_{\langle ii' \rangle, a}  \frac{K (1-\lambda^2)}{1+\lambda^2} \tilde{t}_{ia}^{\dagger} \tilde{t}_{i'a}
+ \sum_{\langle ii' \rangle, a} \frac{K}{2} (\tilde{t}_{ia}^{\dagger} \tilde{t}_{i'a}^{\dagger} + h.c. ) \nonumber\\
&+\sum_{i} \left[ J\frac{1-\lambda^2}{1+\lambda^2} - \frac{4 \lambda \hz}{1+\lambda^2} +
\frac{8Kd \lambda^2}{(1+\lambda^2)^2} \right] \tilde{t}_{iz}^{\dagger} \tilde{t}_{iz} 
+ \sum_{\langle ii' \rangle} \frac{K (1-\lambda^2)^2}{2(1+\lambda^2)^2}
(\tilde{t}_{iz}^{\dagger} \tilde{t}_{i'z}^{\dagger} + \tilde{t}_{iz}^{\dagger} \tilde{t}_{i'z} + h.c.) \,,
\end{align}
\begin{align}
\label{H3}
\mathcal{H}_3 &= \frac{2K}{1+\lm^2} \sum_{\langle ii' \rangle}
\lm_{i} \left[ \tilde{t}_{ix}^{\dagger} \tilde{t}_{i'z}^{\dagger} \tilde{t}_{i'x}
+ \tilde{t}_{iy}^{\dagger} \tilde{t}_{i'z}^{\dagger} \tilde{t}_{i'y} + h.c. \right] 
-\frac{2 K (1-\lm^2)}{(1+\lambda^2)^2} \sum_{\langle ii' \rangle} \lm_{i'} \left[ 
\sum_{\gamma} \tilde{t}_{iz}^{\dagger} \tilde{t}_{i'\gamma}^{\dagger} \tilde{t}_{i'\gamma}
+  \tilde{t}_{iz}^{\dagger} \tilde{t}_{i'z}^{\dagger} \tilde{t}_{i'z}
+ h.c. \right] \nonumber\\
&+ \left[ \frac{2 K\lambda (1-\lambda^2)}{(1+\lambda^2)^2}- \frac{J \lambda}{1+\lambda^2} - \frac{\hz (1-\lambda^{2})}{1+\lambda^{2}}\right]\sum_{i,\gamma} e^{i \vec{Q} \cdot \vec{r_i}}
\left[ \tilde{t}_{iz}^{\dagger} \tilde{t}_{i\gamma}^{\dagger} \tilde{t}_{i\gamma} + h.c. \right]
\,,
\\
\label{H4}
\mathcal{H}_4 &= -\frac{K}{2(1+\lambda^2)} \sum_{\langle ii' \rangle, a}  \left[ 2 \sum_{\gamma}  \right.
\left[(1+\lambda^2) \tilde{t}_{ia}^{\dagger} \tilde{t}_{i'a}^{\dagger} \tilde{t}_{i'\gamma}^{\dagger} \tilde{t}_{i'\gamma} +
(1-\lambda^2) \tilde{t}_{ia}^{\dagger} \tilde{t}_{i'\gamma}^{\dagger} \tilde{t}_{i'\gamma} \tilde{t}_{i'a} \right] \nonumber \\
&+ (1+\lambda^2) \tilde{t}_{ia}^{\dagger} \tilde{t}_{i'a}^{\dagger} \tilde{t}_{iz} \tilde{t}_{i'z} 
\left.- (1-\lambda^2) \tilde{t}_{ia}^{\dagger} \tilde{t}_{i'z}^{\dagger} \tilde{t}_{iz} \tilde{t}_{i'a} + h.c.  \right] \nonumber\\
&- \frac{K}{2(1+\lambda^2)^2} \sum_{\langle ii' \rangle} \left[ 2\sum_{\gamma} \right.
\left[ (1-\lambda^2)^2 \tilde{t}_{i'z}^{\dagger} \tilde{t}_{iz}^{\dagger} \tilde{t}_{i\gamma}^{\dagger} \tilde{t}_{i\gamma} +
(1-\lambda^2)^2 \tilde{t}_{iz}^{\dagger} \tilde{t}_{i\gamma}^{\dagger} \tilde{t}_{i\gamma} \tilde{t}_{i'z} +
2 \lambda^2 \tilde{t}_{iz}^{\dagger} \tilde{t}_{i'\gamma}^{\dagger} \tilde{t}_{i'\gamma} \tilde{t}_{iz} \right] \nonumber \\
&+2 \sum_{\gamma,\delta}\lambda^2 \tilde{t}_{i\gamma}^{\dagger} \tilde{t}_{i'\delta}^{\dagger} \tilde{t}_{i\gamma} \tilde{t}_{i'\delta} 
+2 \lambda^2 \tilde{t}_{iz}^{\dagger} \tilde{t}_{i'z}^{\dagger} \tilde{t}_{iz} \tilde{t}_{i'z} +
(1+\lambda^2)^2 \tilde{t}_{ix}^{\dagger} \tilde{t}_{i'x}^{\dagger} \tilde{t}_{iy} \tilde{t}_{i'y}
\left.- (1+\lambda^2)^2 \tilde{t}_{ix}^{\dagger} \tilde{t}_{i'y}^{\dagger} \tilde{t}_{iy} \tilde{t}_{i'x} + h.c. \right].
\end{align}
\end{widetext}
Here, summations over $a$ refer to the transverse components $a=x,y$, while $\gamma,\delta=x,y,z$. This reflects the fact that the transverse modes ($x,y$) are degenerate, but distinct from the longitudinal ($z$) one.


\subsection{Strategy for $1/d$ expansion}

As in I, the basis for the $1/d$ expansion is the observation that a suitably chosen product state $|\tilde{\psi}_0\rangle$ delivers exact expectation values of local observables in the limit $d\to\infty$, with corrections vanishing as $1/d$.
While in the paramagnetic phase this reference state is simply spanned by local singlets, the triplet admixture parameterized by $\lm$ in Eq.~\eqref{rotate} will vary as function of the coupling ratio $\qq$ inside the AF phase, such that the condensate parameter $\lm$ acquires a $1/d$ expansion. As will be shown below, $\lm$ is proportional to the staggered magnetization (at small $\lm$), such that $\lm$ is expected to vary in a non-analytic, but mean-field-like, fashion near the
QPT. According to the discussion in Section~\ref{sec:mfchar}, this suggests to expand $\lm^2$ in a Taylor series in $1/d$ via the following ansatz:
\begin{align}
\label{l}
\lm^2 = \lm_{0}^{2} + \frac{\lm_{1}}{d} + \frac{\lm_{2}}{d^2} + \ldots
\end{align}

The $1/d$ expansion now requires to perform perturbation theory in the non-linear couplings of $\mathcal{H}$ and, at the same time, to determine the corrections to $\lambda$, keeping in mind that the Hamiltonian itself formally depends on $\lambda$.


\subsection{Linear part}

The condensate parameter $\lambda$ must be chosen such that Hamiltonian pieces which are {\em linear}
in $\tilde{t}$ operators vanish, because these pieces would generate an additional condensate.
To leading order, this translates into $\mathcal{H}_1=0$, i.e.
\begin{equation}
\label{h1a}
h_{1a} (\lm,\hz) \equiv \frac{\lambda J}{1+\lambda^2} - \frac{2\qq J \lambda (1-\lambda^2)}{(1+\lambda^2)^2} +
\frac{\hz (1-\lambda^2)}{1+\lambda^2} = 0 \,.
\end{equation}
We denote the solution of this equation by $\lm_{0}(\hz)$; for $\hz = 0$ it reads
\begin{equation}
\label{lm0}
\lm_{0}^{2}(\hz\!=\!0) = \frac{2\qq-1}{2\qq+1} \,.
\end{equation}
The same result can be obtained variationally by minimizing $\langle{\tilde\psi}_0|\mathcal{H}|{\tilde\psi}_0\rangle$.

From Eq.~\eqref{lm0} we have, on the one hand, $|\lm_0|\to1$ for $\qq\to\infty$, i.e., a classical N\'{e}el state emerges as the reference state in the limit of decoupled ``layers''. On the other hand, $\lm\to0$ as $\qq\to1/2^+$: The ordered state ceases to exist at the quantum critical point at $\qqc=1/2$. This coincides with the leading-order result for the phase boundary obtained in I. Corrections to $\lm$ according to Eq.~\eqref{l} will yield $1/d$ corrections to the phase boundary.
Finally, we note that a dominant staggered field, $|\hz|\gg J,qJ$, also results in $|\lm_0|\to1$.


\subsection{Harmonic approximation}
\label{sec:harm}

The bilinear part of the $\tilde{t}$ Hamiltonian, $\mathcal{H}_2$ in Eq.~\eqref{H2}, takes the following form in momentum space:
\begin{equation}
\label{H2k}
\mathcal{H}_2(\lambda) =  \sum_{\kk,\alpha}
\left[ A_{\vec{k}\alpha} \tilde{t}_{\vec{k} \alpha}^{\dagger} \tilde{t}_{\vec{k} \alpha} + \right.
\left. \frac{B_{\vec{k}\alpha}}{2} (\tilde{t}_{\vec{k} \alpha}^{\dagger} \tilde{t}_{-\vec{k} \alpha}^{\dagger} + h.c.) \right].
\end{equation}
Here, momenta $\vec{k}$ are taken from the {\em full} first Brillouin zone, and
the $\lambda$-dependent coefficients read:
\begin{align}
\label{ax}
A_{\kk a} &= \frac{J}{1+\lambda^2} - \frac{2\lambda \hz}{1+\lambda^2} +
\frac{4\qq J\lm^2}{(1+\lm^2)^2} + \frac{1-\lm^2}{1+\lm^2} B_{\kk a} \,,\\
\label{bx}
B_{\kk a} &= \qq J \gk\,, \\
\label{az}
A_{\kk z} &= J \frac{1-\lambda^2}{1+\lambda^2} - \frac{4\lambda \hz}{1+\lambda^2} +
\frac{8\qq J\lm^2}{(1+\lm^2)^2} +  B_{\kk z}\,, \\           
\label{bz}
B_{\kk z} &= \qq J \gk \left(\frac{1-\lm^2 }{1+\lm^2}\right)^{2}
\end{align}
where $\gk$ is the normalized interaction structure factor
\begin{equation}
\gk = \frac{1}{d} \sum_{n=1}^d \cos k_n\,.
\label{gammadef}
\end{equation}
Notably, there is no mixing between the three excitation modes at the harmonic level -- this is specific to the present case of collinear order [and to the basis choice in Eqs.~(\ref{rotate2},\ref{rotate3})] and would not apply to excitations of canted states.\cite{sommer}

To set the stage for a perturbative treatment, we define the leading (in $1/d$) piece of this bilinear Hamiltonian as unperturbed system, $\mathcal{H}_2^{(0)} \equiv \mathcal{H}_2(\lm_0)$.
Its coefficients are $A_{\vec{k}\alpha}^{(0)} \equiv A_{\vec{k}\alpha}(\lm_0)$ and $B_{\vec{k}\alpha}^{(0)} \equiv B_{\vec{k}\alpha}(\lm_0)$;
using $\hz(\lm_0)$ from Eq.~\eqref{h1a} the $A_{\vec{k}\alpha}^{(0)}$ can be brought in the form
\begin{equation}
\label{abxy}
A^{(0)}_{\kk a} = \A +  \frac{1-\lm_{0}^{2}}{1+\lm_{0}^{2}} B^{(0)}_{\kk a} \,,~~
A^{(0)}_{\kk z} = \Ap +  B^{(0)}_{\kk z} \,
\end{equation}
with the shorthands
\begin{equation}
\label{a}
\A = \frac{J}{1-\lm_{0}^{2}}, ~~
\Ap= J \frac{1+\lm_{0}^{2}}{1-\lm_{0}^{2}} \,.
\end{equation}

The solution of $\mathcal{H}_2^{(0)}$ can be obtained by a standard Bogoliubov transformation,
\begin{equation}
\label{bogol}
\tilde{t}_{\vec{k}\alpha} =
\ut_{\vec{k}}\tb_{\vec{k}\alpha}+\vt_{\vec{k}}\tb^\dagger_{-\vec{k},\alpha},
\end{equation}
and will be dubbed ``harmonic approximation''.
The Bogoliubov coefficients obey
\begin{align}
\label{eq3}
\ut_{\kk \alpha}^2, \vt_{\kk \alpha}^2 &= \pm \frac{1}{2} +
\frac{A^{(0)}_{\kk \alpha}}{2\wt_{\kk \alpha}}\,,~~
\ut_{\kk \alpha} \vt_{\kk \alpha} = - \frac{B^{(0)}_{\kk \alpha}}{2\wt_{\kk \alpha}} \,,
\end{align}
with the eigenmode energies
\begin{equation}
\wt_{\kk \alpha} = \sqrt{ {A^{(0)}_{\kk\alpha}}^{2} - {B^{(0)}_{\kk\alpha}}^{2}}\,.
\end{equation}

While the above formulas are valid for arbitrary staggered field $\hz$, we can obtain explicit expressions for the case $\hz=0$ using Eq.~\eqref{lm0}:
\begin{equation}
\label{a0}
\A= \frac{(2\qq +1)J}{2} \,,~~~~ \Ap= 2\qq J \,, \\
\end{equation}
leading to
\begin{align}
\label{wa0}
\wt_{\kk a} &= J\frac{2\qq +1}{2} \sqrt{1 + \frac{2\gk}{2\qq+1} - \frac{2\qq-1}{2\qq+1}\gk^2} \,, \\
\label{wz0}
\wt_{\kk z} &= 2J\qq \sqrt{1 + \frac{\gk}{4\qq^2}} \,.
\end{align}
A discussion of the dispersions is deferred to Section~\ref{sec:disp} below.


\subsection{Normal-ordered Hamiltonian}

To apply diagrammatic perturbation theory, we need to convert the Hamiltonian into a normal-ordered form in terms of bosons which diagonalize the free-particle piece.
As in I, we employ the strategy to Bogoliubov-transform the leading-order bilinear terms only, according to Eqs.~\eqref{bogol} and \eqref{eq3}. Consequently, additional bilinear terms, obtained both from corrections to the condensate parameter $\lm$ and from normal ordering of higher-order terms, need to be treated perturbatively.

After expressing the Hamiltonian via the $\tb$ operators and subsequent normal ordering, it takes the form
\begin{equation}
\label{hpexp}
\mathcal{H} = \mathcal{H}'_0 + \mathcal{H}'_1 + \mathcal{H}'_2 + \mathcal{H}'_3 + \mathcal{H}'_4 + \mathcal{H}'_5 + \mathcal{H}'_6
\end{equation}
where $\mathcal{H}'_n(\lambda)$ contains $n$ of the Bogoliubov-transformed $\tb$ operators.
The $1/d$ expansion of $\lm$ \eqref{l} can be used to formally split each $\mathcal{H}'_n$ into pieces arising from the different orders in the $\lm$ expansion:
\begin{align}
\label{Hl}
\mathcal{H}'_n(\lambda) &= {\mathcal{H}'_n}^{(0)} +  {\mathcal{H}'_n}^{(1)} + {\mathcal{H}'_n}^{(2)} + \ldots
\end{align}
where ${\mathcal{H}'_n}^{(0)} \equiv \mathcal{H}'_n(\lambda_0)$, ${\mathcal{H}'_n}^{(1)} = \mathcal{H}'_n(\sqrt{\lambda_0^2+\lambda_1/d}) - \mathcal{H}'_n(\lambda_0)$ and so on.
With this prescription, all terms in a particular piece ${\mathcal{H}'_n}^{(m)}$ are at least suppressed as $1/d^m$. We will make frequent use of this splitting in the course of evaluating observables in the next Section.

We will now quote selected pieces of the $\tb$ Hamiltonian which are needed for the following calculations. As above, we restrict ourselves to terms arising from $\mathcal{H}_{0,\ldots,4}$, as these are sufficient to obtain the desired corrections to the order parameter and to the mode dispersion to order $1/d$.
The constant term is
\begin{align}
\label{h0p}
\mathcal{H}'_0 &= -\frac{N J (3-\lambda^2)}{4(1+\lambda^2)} -
\frac{2NKd\lambda^2}{(1+\lambda^2)^2} + \frac{2 N \hz \lambda}{1+\lambda^2} \nonumber \\
&+ \sum_{\kk,\alpha} \left[ A_{\kk\alpha} \vt_{\kk\alpha}^{2} + B_{\kk\alpha} \ut_{\kk\alpha} \vt_{\kk\alpha} \right] + \ldots
\end{align}
where the second line arises from normal ordering of $\mathcal{H}_2$. Additional terms from normal ordering of $\mathcal{H}_4$ are of order $1/d^2$ and are {\em not} shown, see Section~\ref{sec:e0} below for further comments.

It is useful to split the bilinear $\tb$ terms into $\mathcal{H}'_2 =  \mathcal{H}'_{2a} + \mathcal{H}'_{2b} + \mathcal{H}'_{2c}$, where $\mathcal{H}'_{2a}$ is the unperturbed (or harmonic) piece,
\begin{equation}
\label{hp2a}
\mathcal{H}'_{2a} = \sum_{\kk,\alpha} \wt_{\kk\alpha} \tb_{\kk\alpha}^{\dagger} \tb_{\kk\alpha},
\end{equation}
while $\mathcal{H}'_{2b}$ contains the remaining terms coming from $\mathcal{H}_2$:
\begin{align}
\mathcal{H}'_{2b} &=
\sum_{\kk,\alpha} \Big\{\Big[ A_{\kk\alpha}^{(r)} (\ut_{\kk\alpha}^{2}\!+\!\vt_{\kk\alpha}^{2}) + 2B_{\kk\alpha}^{(r)} \ut_{\kk\alpha} \vt_{\kk\alpha} \Big]
\tb_{\kk\alpha}^{\dagger} \tb_{\kk\alpha} \nonumber \\
&+ \Big[ A_{\kk\alpha}^{(r)} \ut_{\kk\alpha}\vt_{\kk\alpha} + \frac{B_{\kk\alpha}^{(r)}}{2} (\ut_{\kk\alpha}^{2}\!+\!\vt_{\kk\alpha}^{2})\Big]
(\tb_{\kk\alpha}^{\dagger} \tb_{-\kk\alpha}^{\dagger} + h.c. )\Big\}
\end{align}
where $A_{\kk\alpha}^{(r)} = A_{\kk\alpha}(\lm) - A_{\kk\alpha}^{(0)}$ and $B_{\kk\alpha}^{(r)} = B_{\kk\alpha}(\lm) - B_{\kk\alpha}^{(0)}$. Finally,
\begin{equation}
\label{hp2b}
\mathcal{H}'_{2c} =
\sum_{\kk\alpha}
\left[
C_{\kk\alpha} \tb^\dagger_{\kk\alpha} \tb_{\kk\alpha} +
\frac{D_{\kk\alpha}}{2} (\tb^\dagger_{\kk\alpha} \tb^\dagger_{-\kk\alpha} + h.c.)
\right]
\end{equation}
represents the bilinear terms generated from normal ordering of $\mathcal{H}_4$, with the coefficients $C_{\kk\alpha}$ and $D_{\kk\alpha}$ listed in Appendix~\ref{app:vert}. Importantly, all contributions to the coefficients in $\mathcal{H}'_{2b}$ and $\mathcal{H}'_{2c}$ are of order $1/d$ or smaller.

The linear-in-$\tb$ piece of the Hamiltonian reads:
\begin{align}
\mathcal{H}'_1 &= \mathcal{H}'_{1a} + \mathcal{H}'_{1b}
= (h_{1a} + h_{1b})(\ut_{\vec{Q}z}+\vt_{\vec{Q}z}) (\tb_{\vec{Q}z}^{\dagger} + \tb_{\vec{Q}z})
\label{h1p}
\end{align}
with $h_{1a}(\lm,\hz)$ from Eq.~\eqref{h1a} and the following contribution from normal ordering of $\mathcal{H}_3$:
\begin{align}
\label{h1b}
h_{1b}= -2 \Ja R_{4a} &+ 2\Jb (R_{2a} + R_{2z} - R_{4z} - R_{3z}) \notag\\
&- h_{1a} (2R_{2a} +2R_{2z} +R_{1z})
\end{align}
with the shorthands
\begin{equation}
\label{j1j2}
\Ja = \frac{2\qq J\lm}{1+\lm^{2}} \,; ~~~\Jb = \Ja \frac{1-\lm^{2}}{1+\lm^{2}}\,.
\end{equation}
The $R_{1\ldots4}$ represent momentum summations over combinations of Bogoliubov coefficients and are listed in Appendix~\ref{app:sum}.

The cubic term involves interactions between a longitudinal and two transverse excitations as well as those of three longitudinal ones. It reads:
\begin{widetext}
\begin{align}
\label{h3p}
\mathcal{H}'_3 &= \sum_{123,a} \left[ \vcxa (\tb_{1a}^{\dagger} \tb_{2z}^{\dagger} \tb_{3a}^{\dagger} + \tb_{1a} \tb_{2z} \tb_{3a}) \delta_{Q+1+2+3}  \right.
+ \vcxb (\tb_{1a}^{\dagger} \tb_{3a}^{\dagger} \tb_{2z} + \tb_{2z}^{\dagger} \tb_{3a} \tb_{1a}) \delta_{Q+1-2+3}  \nonumber \\
&\left. + \vcxc (\tb_{1a}^{\dagger} \tb_{2z}^{\dagger} \tb_{3a} + \tb_{3a}^{\dagger} \tb_{2z} \tb_{1a}) \delta_{Q+1+2-3} \right]   \nonumber \\
&+\sum_{123} \left[ \vcza (\tb_{1z}^{\dagger} \tb_{2z}^{\dagger} \tb_{3z}^{\dagger} + \tb_{1z} \tb_{2z} \tb_{3z}) \delta_{Q+1+2+3}
+ \vczb (\tb_{1z}^{\dagger} \tb_{2z}^{\dagger} \tb_{3z} + \tb_{3z}^{\dagger} \tb_{2z} \tb_{1z}) \delta_{Q+1+2-3} \right],
\end{align}
where the $\delta$ functions account for momentum conservation up to reciprocal lattice vectors of the hypercubic lattice, and their arguments reflect the fact that the condensate is staggered, i.e., each longitudinal ($\tb_{z}$) excitation carries an additional momentum ${\vec Q}$.
Finally, the normal-ordered quartic term may be split as $\mathcal{H}'_4 = \mathcal{H}'^{az}_{4} + \mathcal{H}'^{z}_{4} + \mathcal{H}'^{ab}_{4}$, with its pieces:
\begin{align}
\mathcal{H}'^{az}_{4} &= \sum_{1234, a} \left[ \vqaza (\tb_{1a}^{\dagger} \tb_{2a}^{\dagger} \tb_{3z}^{\dagger} \tb_{4z}^{\dagger} + \tb_{1a}\tb_{2a}\tb_{3z}\tb_{4z})  \right.
\delta_{1+2+3+4}  \nonumber \\
&+ (\vqazb\tb_{1a}^{\dagger}\tb_{2a}^{\dagger}\tb_{3z}\tb_{4z} + \vqazc\tb_{1a}^{\dagger}\tb_{2z}^{\dagger}\tb_{3a}\tb_{4z}
+ \vqazd\tb_{1z}^{\dagger}\tb_{2z}^{\dagger}\tb_{3a}\tb_{4a} ) \delta_{1+2-3-4}   \nonumber \\
&\left. + \vqaze (\tb_{1a}^{\dagger}\tb_{2a}^{\dagger}\tb_{3z}^{\dagger}\tb_{4z} + \tb_{4z}^{\dagger}\tb_{3z}\tb_{2a}\tb_{1a}) \delta_{1+2+3-4}
+ \vqazf(\tb_{1z}^{\dagger}\tb_{2z}^{\dagger}\tb_{3a}^{\dagger}\tb_{4a} + \tb_{4a}^{\dagger}\tb_{3a}\tb_{2z}\tb_{1z}) \delta_{1+2+3-4} \right], \\
\mathcal{H}'^{z}_{4} &= \sum_{1234} \left[ \vqza (\tb_{1z}^{\dagger}\tb_{2z}^{\dagger}\tb_{3z}^{\dagger}\tb_{4z}^{\dagger} + \tb_{1z}\tb_{2z}\tb_{3z}\tb_{4z}) \delta_{1+2+3+4} \right.
+ \vqzb \tb_{1z}^{\dagger}\tb_{2z}^{\dagger}\tb_{3z}\tb_{4z} \delta_{1+2-3-4} \nonumber \\
&\left. + \vqzc (\tb_{1z}^{\dagger}\tb_{2z}^{\dagger}\tb_{3z}^{\dagger}\tb_{4z} + \tb_{4z}^{\dagger}\tb_{3z}\tb_{2z}\tb_{1z}) \delta_{1+2+3-4} \right], \\
\mathcal{H}'^{ab}_{4} &= \sum_{1234, ab} \left[ \vqaba (\tb_{1a}^{\dagger}\tb_{2a}^{\dagger}\tb_{3b}^{\dagger}\tb_{4b}^{\dagger}  \right.
+ \tb_{1a}\tb_{2a}\tb_{3b}\tb_{4b}) \delta_{1+2+3+4}
+ (\vqabb \tb_{1a}^{\dagger}\tb_{2a}^{\dagger}\tb_{3b}\tb_{4b} + \vqabc \tb_{1a}^{\dagger}\tb_{2b}^{\dagger}\tb_{3a}\tb_{4b}) \delta_{1+2-3-4} \nonumber \\
&\left. + \vqabd (\tb_{1a}^{\dagger}\tb_{2a}^{\dagger}\tb_{3b}^{\dagger}\tb_{4b} + \tb_{4b}^{\dagger}\tb_{3b}\tb_{2a}\tb_{1a}) \delta_{1+2+3-4} \right].
\end{align}
\end{widetext}
Explicit expressions for selected vertex functions $\Phi_{3,4}$ are given in Appendix~\ref{app:vert}.


\section{$1/d$ expansion for observables}
\label{sec:expa}

With the Hamiltonian at hand, we are now ready to evaluate observables in the collinear phase of the hypercubic coupled-dimer model in an expansion in $1/d$. As in I, the calculation will be restricted to the leading $1/d$ corrections beyond the harmonic approximation.

The first step is to ensure that the linear-in-$\tb$ piece $\mathcal{H}'_1$ vanishes. Subsequently, standard diagrammatic perturbation theory will be applied, with $\mathcal{H}'_{2a}$ as unperturbed piece and $\mathcal{H}'_{2b}+\mathcal{H}'_{2c}+\mathcal{H}'_{3}+\mathcal{H}'_{4}+\mathcal{H}'_{5}+\mathcal{H}'_{6}$ as perturbation. We exclusively consider zero temperature, where all Hartree loops of $\tb$ particles vanish.

\subsection{Reference product state and phase boundary}

The condition of having no condensate-generating piece in the final Hamiltonian, $\mathcal{H}'_1=0$, can be used to generate a $1/d$ expansion for the condensate parameter $\lm$. To cover the quantum critical point, the expansion needs to be done for $\lm^2$, with the parametrization as in Eq.~\eqref{l}.

\begin{figure}[tb]
\includegraphics[width=0.47\textwidth]{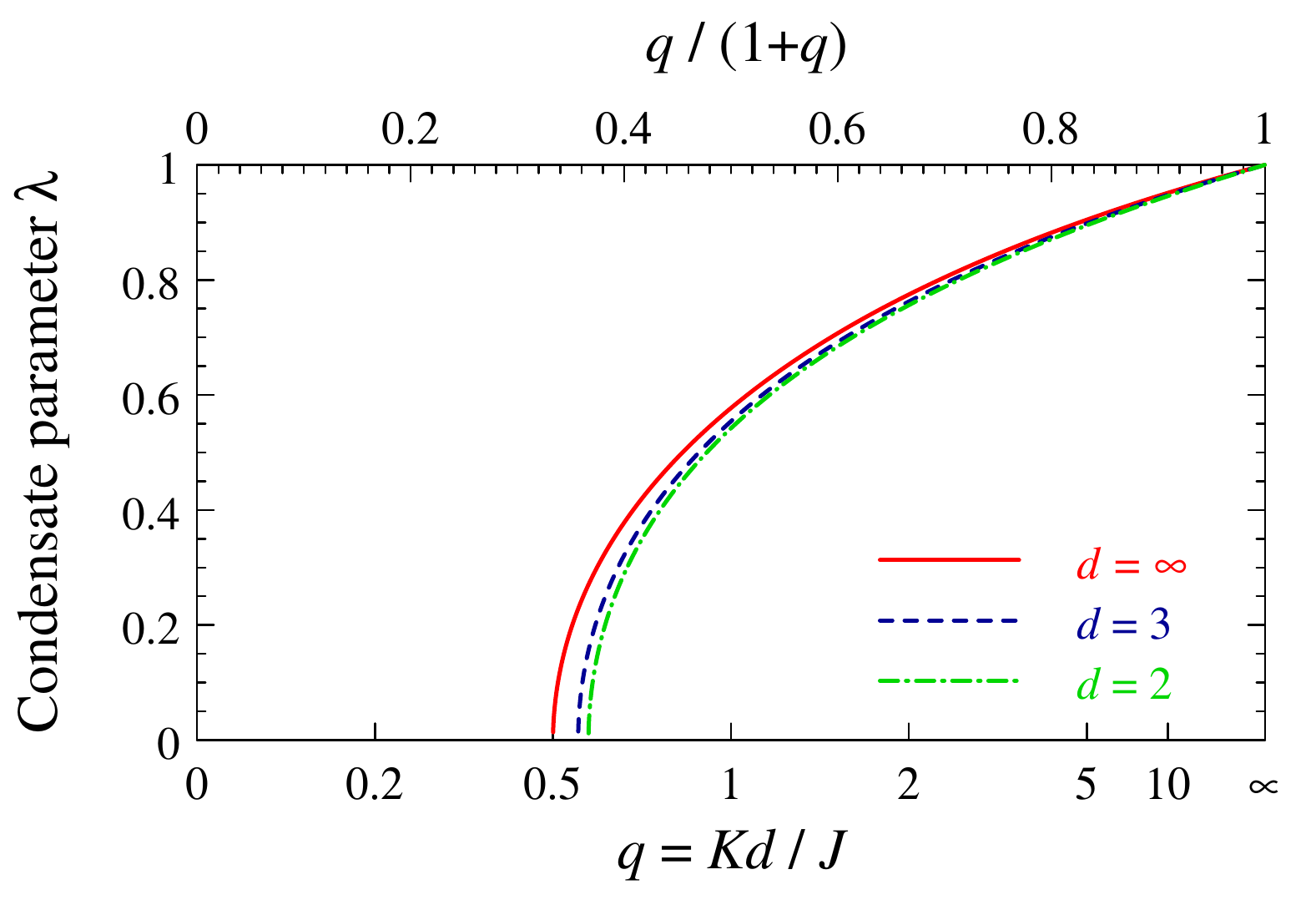}
\caption{
Condensate parameter \eqref{lfull} for the coupled-dimer model \eqref{hh} at $\hz=0$. The curves correspond to $d=\infty$ (solid), $d=3$ (dashed), and $d=2$ (dash-dot).
Note that $q/(1+q)=Kd/(J+Kd)$ varies linearly along the horizontal axis.
}
\label{fig:lam}
\end{figure}

From the explicit form of $\mathcal{H}'_1$ \eqref{h1p} we read off the condition $h_{1a}+h_{1b}=0$. Recalling that $h_{1a}(\lm_0,\hz)=0$, we see that the $1/d$ corrections arise from $h_{1a}(\lm\!-\!\lm_0,\hz)$ and $h_{1b}(\lm,\hz)$. The latter can be evaluated at $\lm_0$, because the $R_{1\ldots4}$ factors in Eq.~\eqref{h1b} are of order $1/d$ or smaller.
Expanding $h_{1a}$ around $\lm_0$ yields to order $1/d$:
\begin{equation}
h_{1a} = \frac{\lm_1}{d} \left[ \frac{J(1-\lm_{0}^{2})}{2\lm_0 (1+\lm_{0}^{2})^{2}} - \frac{\qq J (1+\lm_{0}^{4}-6\lm_{0}^{2})}{\lm_0 (1+\lm_{0}^{2})^{3}}
- \frac{2 \hz}{(1+\lm_{0}^{2})^{2}}\right]
\end{equation}
which has to equal $-h_{1b}$.
Using $\hz(\lm_0)$ from Eq.~\eqref{h1a} and solving for $\lm_1$ we find
\begin{equation}
\label{l1}
\frac{\lm_1}{d} = -\frac{4 \Jb \lm_{0} (R_{2a}+R_{2z}-R_{3z}) (1-\lm_{0}^{2})(1+\lm_{0}^{2})^{3}}
{J(1+\lm_{0}^{2})^{3} - 2\qq J (1-\lm_{0}^{2})^{3}},
\end{equation}
where we have used that the $R_{4}$ are of order $1/d^2$ and can be neglected.
This condensate correction can be simplified in the case $\hz=0$ using $\lm_0$ from Eq.~\eqref{lm0}:
\begin{equation}
\frac{\lm_1}{d} = - \frac{8\qq}{(2\qq +1)^{2}} (R_{2a}+R_{2z}-R_{3z}) \,.
\end{equation}
Using the explicit values of $R_{2,3}$ from Appendix~\ref{app:sum}, we thus obtain the following result for the condensate parameter at $\hz=0$:
\begin{align}
\label{lfull}
\lm^{2} = \frac{2\qq-1}{2\qq+1} 
- \frac{1}{d} \left[\frac{4 \qq^{3}}{(2\qq +1)^{4}} + \frac{16 \qq^{2} +1}{64 \qq^{3} (2\qq +1)^{2}}  \right] 
+ \mathcal{O}\left(\frac{1}{d^2}\right),
\end{align}
as illustrated in Fig.~\ref{fig:lam}. For $\qq\to\infty$, there are no fluctuation corrections to $|\lm|=1$: we expect this result to hold to all orders in $1/d$, as $|\lm|\neq 1$ implies entanglement between the ``layers'' which must be absent for $J=0$.

The condition $\lm^2=0$ describes the vanishing of the condensate parameter and can be used to determine the location of the quantum critical point. The ansatz $\qqc = 1/2 + \qq_{1c}/d$ plugged into Eq.~\eqref{lfull} yields the phase boundary of the ordered phase as:
\begin{equation}
\label{qqc}
\qqc = \frac{1}{2} + \frac{3}{16 d} + \mathcal{O}\left(\frac{1}{d^2}\right) \,.
\end{equation}
Importantly, the same expression was obtained in I for the boundary of the disordered phase, by using the condition of a vanishing triplon gap.
Hence, the $1/d$ expansion correctly yields a second-order QPT, with a continuous onset of the order parameter upon increasing $\qq$.

\subsection{Ground-state energy}
\label{sec:e0}

We continue by determining the ground-state energy $\EE$. In the $\tb$-particle formalism, it is given by $\mathcal{H}'_{0}$ \eqref{h0p} plus perturbative corrections from
$\mathcal{H}'_{2b}+\mathcal{H}'_{2c}+\mathcal{H}'_{3}+\mathcal{H}'_{4}+\mathcal{H}'_{5}+\mathcal{H}'_{6}$.
The constant $\mathcal{H}'_{0}$ depends on the condensate parameter $\lm$ and needs to be expanded in $1/d$, using the $1/d$ expansion for $\lm$ itself.
Given that we have determined this expansion to order $1/d$, we can calculate $\EE$ only up to this order -- this is distinct from the disordered-phase calculation in I where we were able to
extract the $1/d^2$ piece as well.
Importantly, the perturbative corrections are of order $1/d^2$ or smaller: The vertices in both $\mathcal{H}'_{2b}$ and $\mathcal{H}'_{2c}$ are of order $1/d$, and the diagrams involving
$\mathcal{H}'_{3,\ldots,6}$ contain at least two momentum summations each contributing at least a factor of $1/d$, for details see I. Hence, we have
\begin{equation}
\EE
= \mathcal{H}'_{0} + \mathcal{O}\left(\frac{1}{d^2}\right)
= \EEz + \frac{\EEo}{d} + \mathcal{O}\left(\frac{1}{d^2}\right).
\end{equation}
where we have parameterized the first two orders in the expansion.

The leading piece $\EEz$ is from $\mathcal{H}_0$ \eqref{H0}, evaluated at $\lm_0$:
\begin{align}
\frac{\EEz}{N} &= -\frac{J (3-\lambda_{0}^{2})}{4 (1+\lambda_{0}^{2})} -
\frac{2 \qq J \lambda_{0}^{2}}{(1+\lambda_{0}^{2})^2} +
\frac{2 \hz \lambda_{0}}{1+\lambda_{0}^{2}} \notag \\
&= -\frac{J(3+\lm_{0}^{2})}{4 (1-\lm_{0}^{2})} + \frac{2 \qq J \lm_{0}^{2}}{(1+\lm_{0}^{2})^{2}},
\label{Eg01}
\end{align}
where $\hz(\lm_0)$ from Eq.~\eqref{h1a} has entered the second equality. $\EEo$ receives contributions from $1/d$ corrections to $\lm$ and from the normal-ordering piece in Eq.~\eqref{h0p},
where the latter can be evaluated at $\lm_0$. The result is
\begin{align}
\frac{\EEo}{N} &= \lm_{1} \left[ \frac{J}{(1+\lm_{0}^{2})^{2}} - 2 \qq J \frac{(1-\lm_{0}^{2})}{(1+\lm_{0}^{2})^{3}} + \frac{\hz (1-\lm_{0}^{2})}{\lm_{0} (1+\lm_{0}^{2})^{2}} \right] \nonumber \\
&+ \frac{ \qq^{2} J^{3}}{8 \Ap^{2}} \left(\frac{1-\lm_{0}^{2}}{1+\lm_{0}^{2}}\right)^{3} - \frac{ \qq^{2} J^{2}}{4\Ap} \left(\frac{1-\lm_{0}^{2}}{1+\lm_{0}^{2}}\right)^{4}  \nonumber \\
&+ \frac{ \qq^{2} J^{3}}{4 \A^{2} (1-\lm_{0}^{2})} - \frac{ \qq^{2} J^{2}}{2 \A},
\end{align}
with $\A$ and $\Ap$ defined in Eq.~\eqref{a}. Eliminating $\hz$ as before and using $\lm_1$ from Eq.~\eqref{l1}, together with the expressions for $R_{2,3}$ from Appendix~\ref{app:sum},
this can be cast into
\begin{align}
\frac{\EEo}{JN} &= -\frac{\qq^{2}}{8} \left(\frac{1-\lm_{0}^{2}}{1+\lm_{0}^{2}}\right)^{5} - \frac{ \qq^{2}}{4} (1-\lm_{0}^{2}).
\label{Eg11}
\end{align}
In the limit $\hz=0$, our final result for the ground-state energy reads
\begin{equation}
\label{e0f}
\frac{\EE}{JN}
= -\frac{4\qq^{2}+2\qq +1}{8\qq} - \frac{1}{2d}\left[\frac{1}{128 \qq^{3}} + \frac{\qq^{2}}{2\qq +1} \right] + \mathcal{O}\left(\frac{1}{d^2}\right).
\end{equation}
This expression is analytic even at the quantum critical point, reflecting the mean-field value\cite{goldenfeld} $\alpha=0$ of the specific-heat exponent $\alpha$.

At the critical point, the above calculation reproduces the ground-state energy obtained in I for the paramagnetic phase. This is most transparent by inserting $\lm=0$ directly into $\mathcal{H}'_0$ from Eq.~\eqref{h0p}, which then yields the leading two terms of the $1/d$ expansion of $E_0$ in the corresponding equation in I. Alternatively, one may set $\lm_0=0$ in Eqs.~\eqref{Eg01} and \eqref{Eg11} to obtain the same result.
A discussion of the limit of vanishing intra-dimer coupling, $\qq\to\infty$, and its connection to spin-wave theory is given in Section~\ref{sec:spinw}.


\subsection{Triplet density}

\begin{figure}[!tb]
\includegraphics[width=0.47\textwidth]{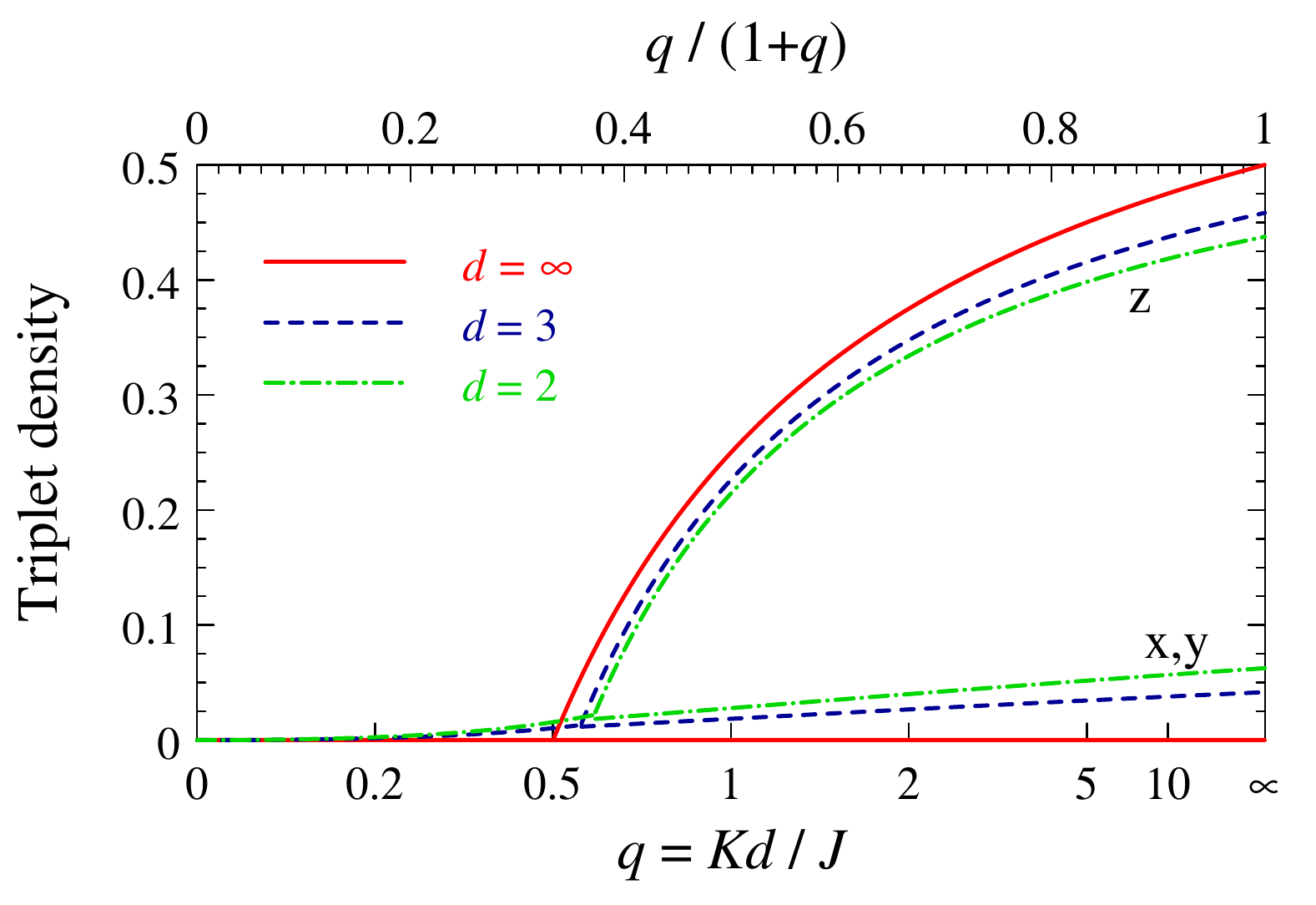}
\caption{
Triplet densities $\langle t_{i\alpha}^{\dagger} t_{i\alpha} \rangle$ for $\alpha=x,y$ and $\alpha=z$
for the hypercubic-lattice dimer model at $\hz=0$ in $d=\infty$ (solid), $d=3$ (dashed), and $d=2$ (dash-dot). In the paramagnetic phase, the densities are given\cite{i} by $\langle t_{i\alpha}^{\dagger} t_{i\alpha} \rangle=\qq^2/(8d)$ to order $1/d$; the result for the antiferromagnetic phase is in Eqs.~(\ref{densx},\ref{densz}).
}
\label{fig:dens}
\end{figure}

Next we calculate the triplet densities, which can be expressed as $\langle t_{i\alpha}^{\dagger} t_{i\alpha} \rangle$ via triplon operators $t$ defined on top of a singlet background,\cite{i} $|t_\alpha\rangle_i = t_{i\alpha}^{\dagger} |t_0\rangle_i$. Using the basis rotation in Eqs.~(\ref{rotate}-\ref{rotate3}) the densities can be expressed in terms of $\tilde{t}$ operators
as follows:
\begin{align}
t_{ia}^{\dagger} t_{ia} &= \tilde{t}_{ia}^{\dagger} \tilde{t}_{ia}~~(a=x,y) \,, \\
t_{iz}^{\dagger} t_{iz} &= \frac{\tilde{t}_{iz}^{\dagger} \tilde{t}_{iz} + \lm^{2} P_{i} + \lm_{i} (\tilde{t}_{iz}^{\dagger}+\tilde{t}_{iz})}{1+\lm^{2}} \,.
\end{align}
For the corresponding expectation values we find to order $1/d$:
\begin{align}
\frac{1}{N} \sum_{i} \langle t_{ia}^{\dagger} t_{ia} \rangle &= \frac{1}{N} \sum_{i} \langle \tilde{t}_{ia}^{\dagger} \tilde{t}_{ia} \rangle = R_{2a} \,, \\
\frac{1}{N} \sum_{i} \langle t_{iz}^{\dagger} t_{iz} \rangle &= \frac{\lm_{0}^{2}}{1+\lm_{0}^{2}} + \frac{\lm_{1}}{d}\frac{1}{(1+\lm_{0}^{2})^{2}} \nonumber \\
&- \frac{2 \lm_{0}^{2}}{1+\lm_{0}^{2}} R_{2a} + \frac{1-\lm_{0}^{2}}{1+\lm_{0}^{2}} R_{2z} \,.
\end{align}
In these expressions, $\sum_{i} \langle \tilde{t}_{i\alpha}^{\dagger} \tilde{t}_{i\alpha} \rangle/N = R_{2\alpha}$ represents the result of the harmonic approximation, with perturbative corrections starting at order $1/d^2$ only.\cite{corrfoot}
Without staggered field, $\hz=0$, we can write the triplet densities as a function of $\qq$:
\begin{align}
\label{densx}
\frac{1}{N} \sum_{i} \langle t_{ia}^{\dagger} t_{ia} \rangle &= \frac{1}{d} \frac{\qq^{2}}{2(2 \qq +1)^{2}} + \mathcal{O}\left(\frac{1}{d^2}\right) \,, \\
\frac{1}{N} \sum_{i} \langle t_{iz}^{\dagger} t_{iz} \rangle &= \frac{2\qq -1}{4\qq} - \frac{1}{d} \left(\frac{\qq^{2}}{2(2\qq +1)^{2}}  + \frac{1}{64 \qq^{3}} \right) \nonumber\\
&+\mathcal{O}\left(\frac{1}{d^2}\right) \,.
\label{densz}
\end{align}
These results are illustrated in Fig.~\ref{fig:dens}, which most prominently shows a kink in the $z$ triplet density at the QPT. Parenthetically, we note that the local spin correlator can be expressed in terms of the triplet densities according to $\vec{S}_{i1}\cdot \vec{S}_{i2} = \sum_{\alpha} t_{i\alpha}^{\dagger} t_{i\alpha} - \frac{3}{4}$.

\subsection{Staggered magnetization}

The staggered magnetization
\begin{equation}
\label{mstdef}
\MM = \frac{1}{N} \sum_{i} e^{i\vec{Q} \cdot R_{i}} \langle S_{i1}^{z} - S_{i2}^{z} \rangle
\end{equation}
represents the order parameter of the collinear antiferromagnet. It is most efficiently determined by taking the derivative of the ground-state energy w.r.t. $\hz$:
\begin{align}
\MM &= \frac{\partial \EE}{N\partial \hz} =
\frac{\partial \EEz}{N\partial \hz} + \frac{1}{d}\frac{\partial \EEo}{N\partial \hz} + \mathcal{O}\left(\frac{1}{d^2}\right).
\end{align}
Given the mean-field value of the order-parameter exponent $\beta=1/2$, we expect $\MM^2$ to vary analytically near the QPT, and consequently we parameterize
\begin{align}
\label{mm}
\MM^2 = \MMz^{2} + \frac{\MMo}{d} + \frac{\MMt}{d^2} + \ldots
\end{align}

Using Eq.~\eqref{Eg01} the leading piece is found as
\begin{equation}
\label{ms0}
\MMz = \frac{2 \lm_{0}}{1+\lm_{0}^{2}}  + \frac{2h_{1a}(\lm_0,\hz)}{1+\lm_0^2} \frac{\partial \lm_{0}}{\partial \hz} = \frac{2 \lm_{0}}{1+\lm_{0}^{2}}\,.
\end{equation}
with $h_{1a}(\lm,\hz)$ in Eq.~\eqref{h1a}. Given that $h_{1a}(\lm_0,\hz)=0$, the second term vanishes -- this also applies to the limit $\lm_0 \rightarrow 0$ where
$\partial \lm_{0}/ \partial \hz$ diverges.
From Eq.~\eqref{Eg11} we have
\begin{align}
\frac{\partial \EEo}{N\partial \hz} &= \frac{5J \qq^{2} \lm_{0}}{2} \frac{(1-\lm_{0}^{2})^{4}}{(1+\lm_{0}^{2})^{6}} \frac{\partial \lm_{0}}{\partial \hz}
+ \frac{J \qq^{2} \lm_{0}}{2} \frac{\partial \lm_{0}}{\partial \hz}  \nonumber \\
&= -\left[ 5 \frac{(1-\lm_{0}^{2})^{4}}{(1+\lm_{0}^{2})^{6}} +1 \right] \label{ms1}\\
&~~~\times \frac{J \qq^{2} \lm_{0} (1-\lm_{0}^{4})}{2\left[ J(1-2\qq) +3\lm_{0}^{2} J(1+2\qq) - 4\hz \lm_{0}^{3}\right]} \,. \nonumber
\end{align}

\begin{figure}[tb]
\includegraphics[width=0.47\textwidth]{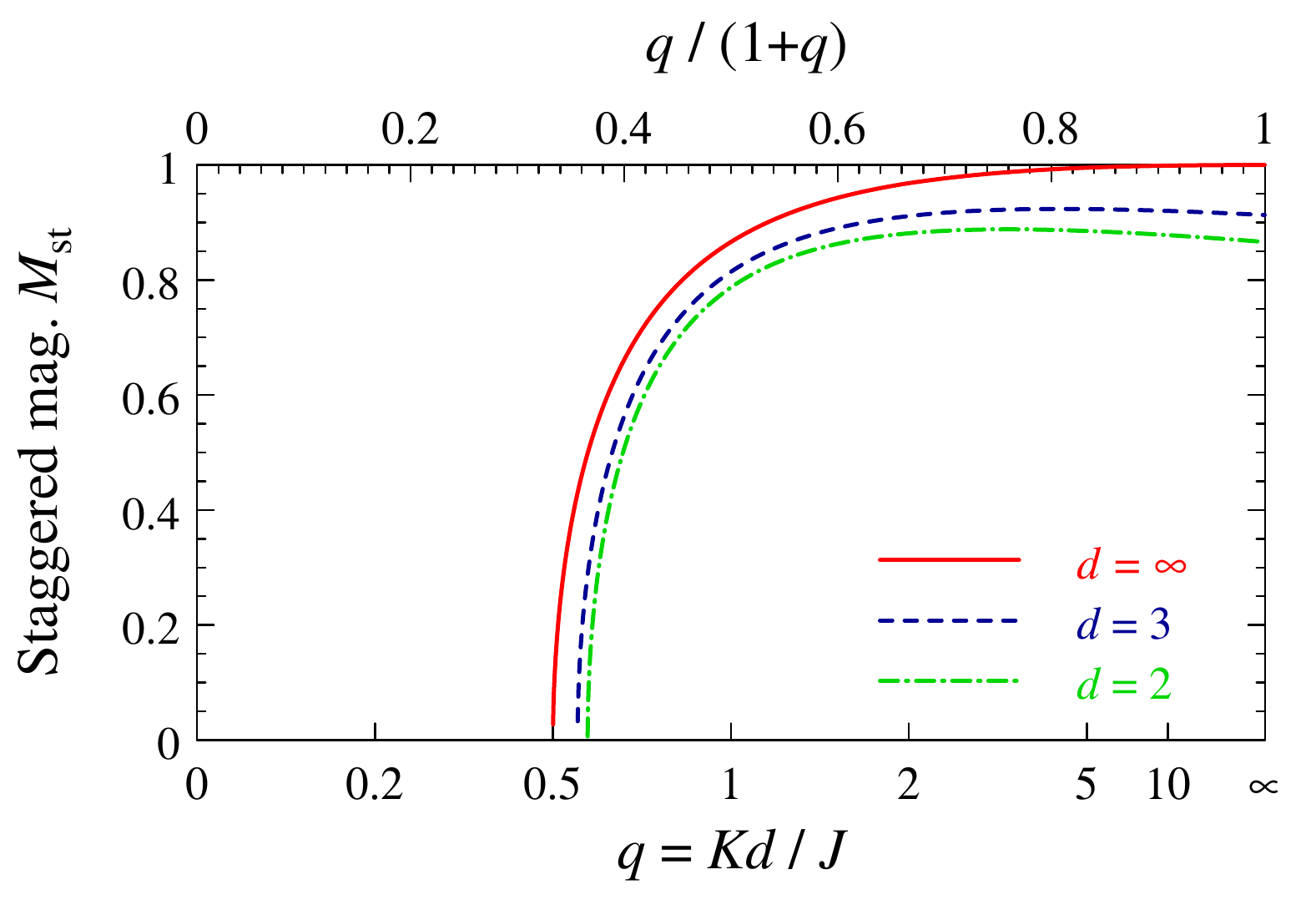}
\caption{
Staggered magnetization per dimer \eqref{msq} derived from the $1/d$ expansion for the coupled-dimer model \eqref{hh} at $\hz=0$. The curves correspond to $d=\infty$ (solid), $d=3$ (dashed), and $d=2$
(dash-dot). Fluctuation corrections lead to a maximum of $\MM$ at some $q<\infty$, see text.
We note that solving $\MM^2=0$ using a truncated series defines a $\qqc(d)$ which is distinct from the expansion result \eqref{qqc} evaluated at fixed finite $d$, because $\MM^2$ from Eq.~\eqref{msq} evaluated at $\qqc$ from Eq.~\eqref{qqc} vanishes only up to order $1/d$ -- this is a natural series-expansion property, as already discussed in Section~IV E of I.
%
}
\label{fig:mst}
\end{figure}


We now focus on the limit $\hz=0$ where Eq.~\eqref{ms1} can be converted into
\begin{equation}
\MMo = - \frac{(1+\lm_{0}^{2})^{2}}{8} \left[ 5 \frac{(1-\lm_{0}^{2})^{4}}{(1+\lm_{0}^{2})^{6}} +1\right],
\end{equation}
representing the second term of the expansion \eqref{mm}.
This yields our final result for the staggered magnetization:
\begin{align}
\label{msq}
\MM^{2} &= \frac{4\qq^{2}-1}{4\qq^{2}} - \frac{1}{d} \frac{2 \qq^{2}}{(2\qq+1)^{2}} \left[ \frac{5 (2\qq+1)^{2}}{256 \qq^{6}} +1 \right] + \mathcal{O}\left(\frac{1}{d^2}\right),
\end{align}
graphically shown in Fig.~\ref{fig:mst}. For $d=\infty$ the magnetization reaches its saturation value in the limit of decoupled ``layers'', $\qq\to\infty$, and fluctuation corrections generically lead to a reduction of $\MM$. Interestingly, these fluctuation effects cause $\MM$ to be maximal at some finite value of the interlayer coupling, indicating that interlayer and intralayer fluctuations compete. This is qualitatively consistent with results for the bilayer square-lattice magnet.\cite{hida92,weihong97}
As shown in Section~\ref{sec:spinw} below, our fluctuation corrections obtained in the limit $\qq\to\infty$ match those obtained from spin-wave theory in this limit.

The vanishing of the order parameter $\MM$ upon decreasing $\qq$ can be used to define the boundary $\qqc$ of the ordered phase, and solving for $\qqc$ we find the same expression as in Eq.~\eqref{qqc},
showing internal consistency of our method.

Last but not least we emphasize that the staggered magnetization cannot only be calculated as a derivative of the ground-state energy, but also directly as the expectation value \eqref{mstdef}, with identical results as required by thermodynamic consistency. Importantly, the expectation-value calculation at order $1/d$ involves both fluctuations around the product state $|\tilde{\psi}_0\rangle$, described by $\tilde{t}$ operators, as well as corrections to $|\tilde{\psi}_0\rangle$, i.e., to the condensate parameter $\lambda$. We note that the latter corrections were overlooked in Ref.~\onlinecite{sommer}; similar problems have appeared in the literature on frustrated hard-core boson systems, see Ref.~\onlinecite{mila12} for a summary.


\subsection{Mode dynamics}
\label{sec:disp}

The elementary excitations of the AF phase are generalized triplons. In contrast to the paramagnetic phase with a triply degenerate excitation spectrum, here we have to distinguish Goldstone and non-Goldstone modes, dubbed transverse and longitudinal, respectively. In the following, we will determine the mode dispersions to order $1/d$, restricting the concrete evaluation to the field-free case $\hz=0$.

The leading-order dispersions are those from the harmonic approximation, $\wt_{\kk a}$ and $\wt_{\kk z}$ displayed in Eqs.~\eqref{wa0} and \eqref{wz0}. Perturbative corrections arise from
$\mathcal{H}'_{2b}+\mathcal{H}'_{2c}+\mathcal{H}'_{3}+\mathcal{H}'_{4}+\mathcal{H}'_{5}+\mathcal{H}'_{6}$ and are suppressed at least as $1/d$. Their calculation parallels that in I, and we refer the reader to that paper for details. In particular, to order $1/d$ it is sufficient to determine the normal self-energies $\Sigma_N$ of the $\tb$ particles, and the renormalized mode energies obey
\begin{equation}
\label{pole1d}
\Wtk^{2} = \wtk^{2} + 2\wtk \Sigma_{N}(\kk,\wtk)\,.
\end{equation}

We first consider the $a=x,y$ modes: as we will see below, these modes remain degenerate and represent the transverse Goldstone modes of the system. The relevant self-energy diagrams contributing to $\mathcal{O}(1/d)$ are listed in Appendix~\ref{app:selfe}, together with their analytic expressions.
Expressing these self-energies at $\hz=0$ in terms of $\qq$ we find, using Eq.~\eqref{pole1d}, the following result for the dispersion of the $a=x,y$ modes:
\begin{widetext}
\begin{align}
\label{dispa1}
\frac{\Wt_{\kk a}^{2}}{J^2} &= \frac{(2\qq+1)^{2}}{4} \left[ 1+\frac{2\gk}{2\qq+1}-\gk^{2} \frac{2\qq-1}{2\qq+1} \right] \nonumber \\
&- \frac{1}{d} \frac{1+\gk}{128 \qq^{2} (2\qq +1)^{2} \left( 32\qq^{2} + 4\qq^{2}\gk^{2} +8\qq-4\qq \gk- \gk^{2} -2\gk \right)}
\left[ 1024\qq^{8} \left(\gk-1\right)\left(8+\gk^{2}\right)  \right. \nonumber \\
&+1024 \qq^{7} \left(\gk-1\right)\left(10-\gk+\gk^{2}\right) -256\qq^{6} \left(\gk+2\right)\left(4+3\gk+\gk^{2}\right) - 256\qq^{5} \left(\gk^{3}-\gk^{2}+20\gk-4\right) \nonumber \\
&- 16\qq^{4} \left(\gk^{3}-21\gk^{2}+64\gk-8 \right)  - 16 \qq^{3} \left(\gk^{3}-8\gk^{2}-29\gk+14\right) +8\qq^{2} \left(-18+25\gk+9\gk^{2}\right)   \nonumber \\
&\left. + 4\qq \left(-10+\gk\left(\gk+1\right)\left(\gk+3\right)\right)  + \left(\gk^{3}+\gk^{2}-2\gk-4\right) \right]
+ \mathcal{O}\left(\frac{1}{d^2}\right).
\end{align}
This expression has the property $\Wt_{\vec{Q}a}^{2} = 0$ for all $\qq$, i.e., both transverse modes are soft at the ordering wavevector. This is the property expected for Goldstone modes; recall that our momenta are taken from the {\em full} first Brillouin zone.
Expanding around $\kk=\vec{Q}$ we can introduce a velocity $c_a$ of the Goldstone mode according to
$\Wt_{\kk a}^2 = c_a^2 (\vec{k}-\vec{Q})^2 / d$, with $c_a$ evaluating to:
\begin{align}
\label{vela}
\frac{c_a}{J} &= \sqrt{\frac{\qq(2\qq+1)}{2}} \left[ 1+ \frac{1}{\qq (2\qq+1)^{3} (6\qq+1) d} \left(12\qq^{5}+14\qq^{4}-2\qq^{3}-4\qq^{2}
-\frac{5\qq}{16}+\frac{13}{32}+\frac{7}{64\qq}+\frac{1}{128\qq^{2}}\right)\right]
+ \mathcal{O}\left(\frac{1}{d^2}\right).
\end{align}
The velocity is non-singular at the QPT, and an explicit evaluation at $\qq=\qqc$, Eq.~\eqref{qqc}, yields Eq.~\eqref{cintro} quoted in Sec.~\ref{sec:introsumm}. Importantly, this velocity equals the longitudinal-mode velocity calculated below, Eq.~\eqref{velz}, as well as the triplon velocity in the disordered phase,\cite{i} if both are evaluated at $\qq=\qqc$. This demonstrates a smooth evolution of the excitation modes across the quantum critical point.

We alert the reader that the connection between the modes discussed here and the signal in inelastic neutron scattering will be discussed in Section~\ref{sec:chi} below. In particular, the distinction between ``even'' and ``odd'' excitations w.r.t. to the spin indices within each dimer will only be made at the level of response functions, while the modes discussed here are defined for dimers and hence do not carry an even/odd quantum number.

We now consider the $z$ mode which will be interpreted as a longitudinal amplitude (or Higgs) mode. The individual contributions to the self-energy are listed in Appendix~\ref{app:selfe}, from which
we obtain the following $1/d$ expansion for the $z$-mode dispersion at $\hz=0$:
\begin{align}
\label{dispz1}
\frac{\Wt_{\kk z}^{2}}{J^2} &=  \left[ 4\qq^{2}+ \gk \right]
+ \frac{1}{32d} \left[ -\frac{1}{\qq^2} - \frac{16}{(2\qq+1)^{2}} + 2\gk^{2} \frac{(\gk-3)(\gk-1)^{2}}{4\qq+1-\gk} \right. \nonumber \\
&-24\frac{(\gk-3)(\gk-1)}{\gk-12\qq^{2}} + 2(4-3\gk+\gk^{2})(-8-\gk+\gk^{2}) + 8 \frac{6-\gk+\gk^{2}}{2\qq+1}
\left.+8\qq(4-\gk^{2}+\gk^{3})\right]
+ \mathcal{O}\left(\frac{1}{d^2}\right).
\end{align}
This dispersion is generally gapped, with a minimum energy at $\kk = \vec{Q}$.
Parameterizing $\Wt_{\kk z}^2 = \Delta_z^2 + c_z^2 (\vec{k}-\vec{Q})^2/d$ we find for
the mode gap:
\begin{equation}
\frac{\Delta_{z}^{2}}{J^2} = 4\qq^{2} -1 + \frac{1}{32d} \left[ -\frac{1}{\qq^2} -\frac{16}{(2\qq+1)^2} + \frac{48}{2\qq+1}
+ \frac{192}{12\qq^2 +1} -96 +16\qq \right]
+ \mathcal{O}\left(\frac{1}{d^2}\right).
\end{equation}
To leading order, we see that $\Delta_z=0$ at $\qq=1/2$. Examining the $1/d$ corrections shows that $\Delta_z=0$ to order $1/d$ for $\qq=\qqc$ from Eq.~\eqref{qqc}, i.e., the gap vanishes at the quantum critical point. In its vicinity, the gap varies with a critical exponent $\nu z = 1/2$ as follows:
\begin{equation}
\frac{\Delta_{z}}{J} = \left[2 - \frac{5}{8d} + \mathcal{O}\left(\frac{1}{d^2}\right)\right] \sqrt{\qq-\qqc}
\end{equation}
which is Eq.~\eqref{delzintro} quoted in the introduction.
The $z$-mode velocity obeys
\begin{equation}
\label{velz}
\frac{c_z}{J} = \frac{1}{\sqrt{2}} \left[ 1+ \frac{1}{32d} \left(\frac{14}{2\qq+1} -\frac{4}{(2\qq+1)^2} - \frac{72}{12\qq^2 +1}
+\frac{96}{(12\qq^2 +1)^2} +6 +20\qq \right) \right]
+ \mathcal{O}\left(\frac{1}{d^2}\right)
\end{equation}
\end{widetext}
which again yields Eq.~\eqref{cintro} if evaluated at $\qq=\qqc$.

Together, this allows us to consistently interpret the $z$ mode as a longitudinal (or Higgs) mode of the ordered phase: It is soft a the QPT where it merges with the gapless transverse modes. Inside the ordered phase, the longitudinal mode is gapped, corresponding to amplitude fluctuations of the AF order parameter. We note that this mode is expected to acquire severe damping deep inside the ordered phase due to two-particle decay into transverse modes.\cite{higgsdamp} However, the corresponding decay rates are exponentially small as $d\to\infty$ and hence cannot be obtained from the $1/d$ expansion.\cite{i}

\begin{figure}[t]
\includegraphics[width=0.47\textwidth]{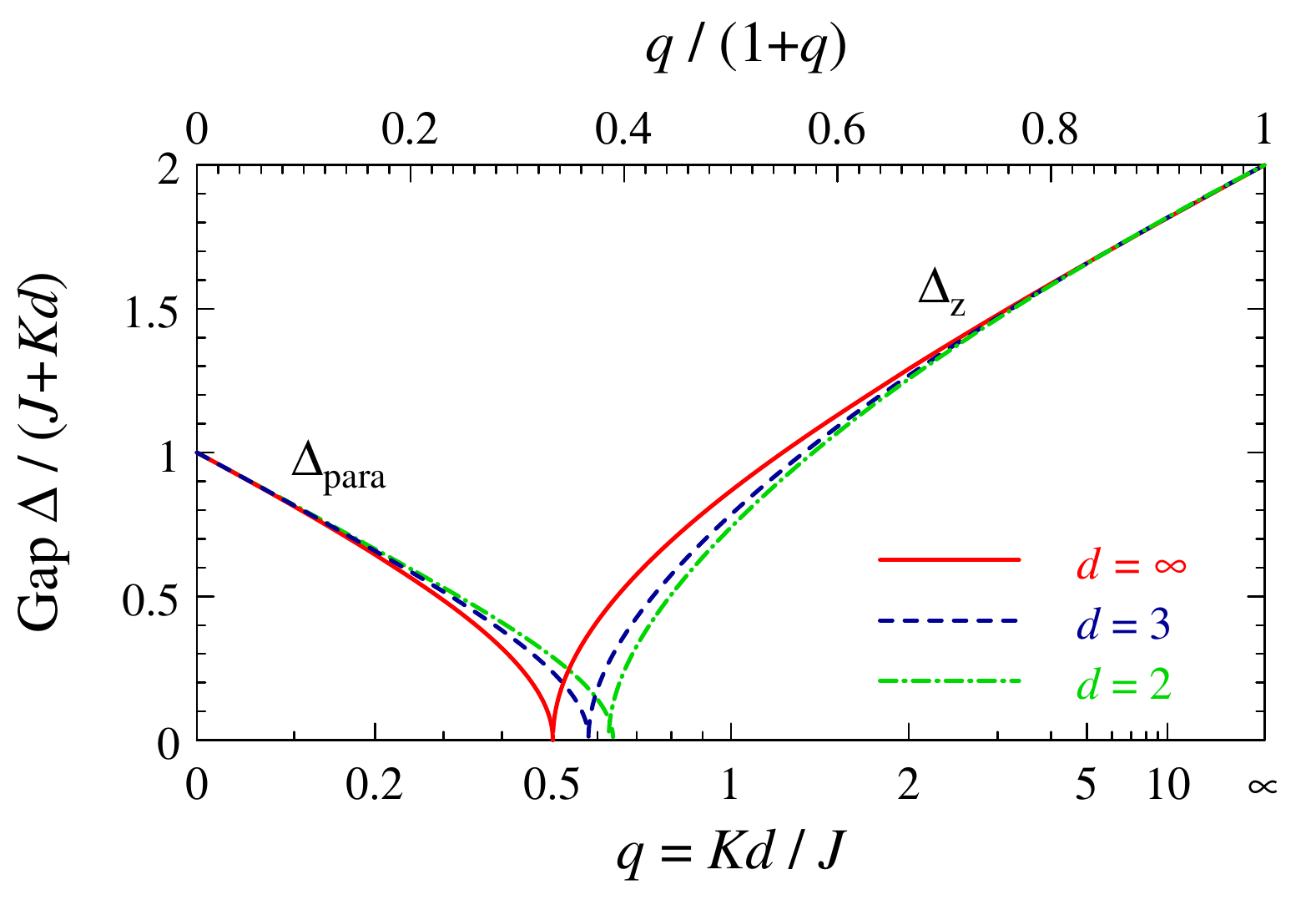}
\caption{
Mode gaps for the hypercubic-lattice coupled dimer model as obtained from the $1/d$ expansion, for $d=\infty$ (solid), $d=3$ (dashed), and $d=2$ (dash-dot).
The triplon gap $\Delta_{\rm para}$ of the paramagnetic phase, as calculated in I, is shown for small $\qq$, whereas the longitudinal (or Higgs) gap $\Delta_z$ is shown for large $\qq$.
Note that $q/(1+q)=Kd/(J+Kd)$ varies linearly along the horizontal axis, and the gaps are plotted as $\Delta/(J+Kd)$.
}
\label{fig:gap}
\end{figure}

Our results for Higgs gap in the ordered phase, combined with those from I for the triplon gap in the disordered phase, are illustrated in Fig.~\ref{fig:gap}. Near criticality we find that their ratio obeys
\begin{equation}
\frac{\Delta_{z} (\qqc + \delta\qq)}{\Delta_{\rm para} (\qqc-\delta\qq)} = \sqrt{2}
\end{equation}
to order $1/d$. In fact, this result has been previously derived\cite{sssolvay} from a $\phi^4$ order-parameter field theory and is valid for any $d$ above the upper critical dimension $d_c^+$. Remarkably, neutron scattering data obtained in Ref.~\onlinecite{ruegg08} for TlCuCl$_3$ have found this relation to be obeyed to good accuracy; for this material $d=d_c^+=3$ such that mean-field behavior is expected up to logarithmic corrections.

\subsection{Dynamic susceptibility}
\label{sec:chi}

We finally connect the excitation modes to the dynamic spin susceptibility,
\begin{equation}
\label{susc}
\chi_{\alpha}(\vec{k},\w) = -\ii \int_{-\infty}^{\infty} dt e^{\ii \w t} \langle T_t S_{\alpha}(\vec{k},t) S_{\alpha}(-\vec{k},0) \rangle,
\end{equation}
as measured by inelastic neutron scattering. For simplicity, we restrict ourselves to the single-mode approximation, i.e., we do not consider excitation continua, and we calculate the distribution of mode weights to leading order $(1/d)^0$ only.

The susceptibility can be probed in the even (e) and odd (o) channel of the each dimer, corresponding to the operators
\begin{equation}
S^{e/o}_{\alpha} = S^{1}_{\alpha} \pm S^{2}_{\alpha}\,.
\end{equation}
These can be re-written using the rotated triplon operators $\tilde{t}$ \eqref{gbond}. The leading-order single-mode expressions read
\begin{align}
S^{e}_{x} (\kk) &= \frac{i \lm}{\sqrt{1+\lm^{2}}} \left[ \tilde{t}_{(\kk-\vec{Q})y}- \tilde{t}^{\dagger}_{(-\kk-\vec{Q})y} \right] \,, \\
S^{e}_{y} (\kk) &= \frac{i \lm}{\sqrt{1+\lm^{2}}} \left[ \tilde{t}^{\dagger}_{(-\kk-\vec{Q})x}- \tilde{t}_{(\kk-\vec{Q})x}\right] \,, \\
S^{e}_{z}(\kk) &= 0 \,, \\
S^{o}_{x} (\kk) &= \frac{\tilde{t}^{\dagger}_{-\kk x} + \tilde{t}_{\kk x}}{\sqrt{1+\lm^{2}}} \,, \\
S^{o}_{y} (\kk) &= \frac{\tilde{t}^{\dagger}_{-\kk y} + \tilde{t}_{\kk y}}{\sqrt{1+\lm^{2}}} \,, \\
S^{o}_{z} (\kk) &= \frac{(1-\lm^{2}) (\tilde{t}^{\dagger}_{-\kk z} + \tilde{t}_{\kk z}) +2\lm \sqrt{N}\delta_{\kk,\vec{Q}} }{1+\lm^{2}} \,.
\label{sozk}
\end{align}
We note that corrections introduced by the projectors $P_i$ \eqref{proj} enter at order $1/d$, and that $S^{e}_{z}$ creates a two-particle continuum only.
Using the Bogoliubov transformation \eqref{bogol} we can express the susceptibility in terms of the $\tb$-Green's functions. Using the fact that $2\vec{Q}$ is a reciprocal lattice vector we obtain:
\begin{widetext}
\begin{align}
 \chi^{e}_{x}(\kk,\w) &= \frac{\lm^{2}}{1+\lm^{2}} \left(u_{(\kk+\vec{Q}) y} - v_{(\kk+\vec{Q}) y}\right)^{2}\left[ \mathcal{G}^{N}_{y}(\kk+\vec{Q},\w) + \mathcal{G}^{N}_{y}(\kk+\vec{Q},-\w)
- \mathcal{G}^{A}_{y}(\kk+\vec{Q},\w) - \mathcal{G}^{A}_{y}(\kk+\vec{Q},-\w)\right] \,,\\
\chi^{o}_{x}(\kk,\w) &= \frac{1}{1+\lm^{2}} (u_{\kk x} + v_{\kk x})^{2}\left[ \mathcal{G}^{N}_{x}(\kk,\w) + \mathcal{G}^{N}_{x}(\kk,-\w)
+ \mathcal{G}^{A}_{x}(\kk,\w) + \mathcal{G}^{A}_{x}(\kk,-\w)\right] \,,
 \\
\chi^{o}_{z}(\kk,\w) &= \left(\frac{1-\lm^{2}}{1+\lm^{2}}\right)^{2} (u_{\kk z} + v_{\kk z})^{2} \left[ \mathcal{G}^{N}_{z}(\kk,\w) + \mathcal{G}^{N}_{z}(\kk,-\w)
+ \mathcal{G}^{A}_{z}(\kk,\w) + \mathcal{G}^{A}_{z}(\kk,-\w)\right]  + \frac{4\lm^{2} N}{(1+\lm^{2})^{2}} \delta(\w) \delta_{\kk,\vec{Q}} \,.
\end{align}
\end{widetext}
The expressions for $\chi_{y}$ are obtained from $\chi_{x}$ by replacing $x \leftrightarrow y$.

To leading order in $1/d$ it is sufficient to evaluate the expressions at the harmonic level, where $\mathcal{G}^{A}=0$ and $\lm = \lm_0$.
Using the degeneracy of the transverse modes, $\wt_{\kk x}=\wt_{\kk y}\equiv \wt_{\kk a}$, $u_{\kk x}=u_{\kk y}\equiv u_{\kk a}$ etc., we obtain for the transverse susceptibilities:
\begin{align}
\chi^{e}_{a}(\kk+\vec{Q},\w) &= \frac{\lm_{0}^{2}(u_{\kk a} - v_{\kk a} )^{2}}{1+\lm_{0}^{2}} \left[ \frac{1}{\w - \wt_{\kk a}}
- \frac{1}{\w + \wt_{\kk a}} \right], \\
\chi^{o}_{a}(\kk,\w) &= \frac{(u_{\kk a} + v_{\kk a})^{2}}{1+\lm_{0}^{2}}  \left[ \frac{1}{\w - \wt_{\kk a}} - \frac{1}{\w + \wt_{\kk a}} \right].
\end{align}
Hence, these susceptibilities obtain single-mode contributions from the transverse modes only. Interestingly, the mode momentum is shifted by $\vec Q$ in the even channel, but unshifted in the odd channel,
and the mode weight in the even channel vanishes upon approaching the transition to the disordered phase. Comparing with the signal in the disordered phase calculated in I, we conclude that the primary signal is
in the odd channel which also smoothly connects to the triplon-mode response of the paramagnet, whereas the signal in the even channel can be interpreted as a replicated signal due to condensate Bragg scattering
(note that the condensate is in the odd channel, i.e., antisymmetric w.r.t. the spin indices in each dimer, as well).

The odd-channel longitudinal susceptibility is
\begin{align}
\chi^{o}_{z}(\kk,\w) &= \left(\frac{1-\lm_{0}^{2}}{1+\lm_{0}^{2}}\right)^{2} \! (u_{\kk z}\!+\!v_{\kk z})^{2} \left[ \frac{1}{\w - \wt_{\kk z}} - \frac{1}{\w + \wt_{\kk z}} \right] \nonumber\\
&+ \frac{4\lm_{0}^{2} N}{(1+\lm_{0}^{2})^{2}} \delta(\w) \delta_{\kk,\vec{Q}} \,,
\end{align}
where the last term $\chi^{o}_{z}$ corresponds to the magnetic Bragg peak of the ordered state, recall $\MM = 2\lm_0 / (1+\lm_0^2)$ to leading order.
We conclude that the amplitude (or Higgs) mode is visible in the longitudinal susceptibility. Upon approaching the QPT, its weight is finite, smoothly connecting to the triplon signal. However, in the limit of vanishing intra-dimer coupling, its weight is zero: In this limit, the mode describes two flipped spins w.r.t. the N\'eel state, see Section~\ref{sec:refstate}, such that it cannot be excited by the action of a single spin operator.

\begin{figure}[b]
\includegraphics[width=0.47\textwidth]{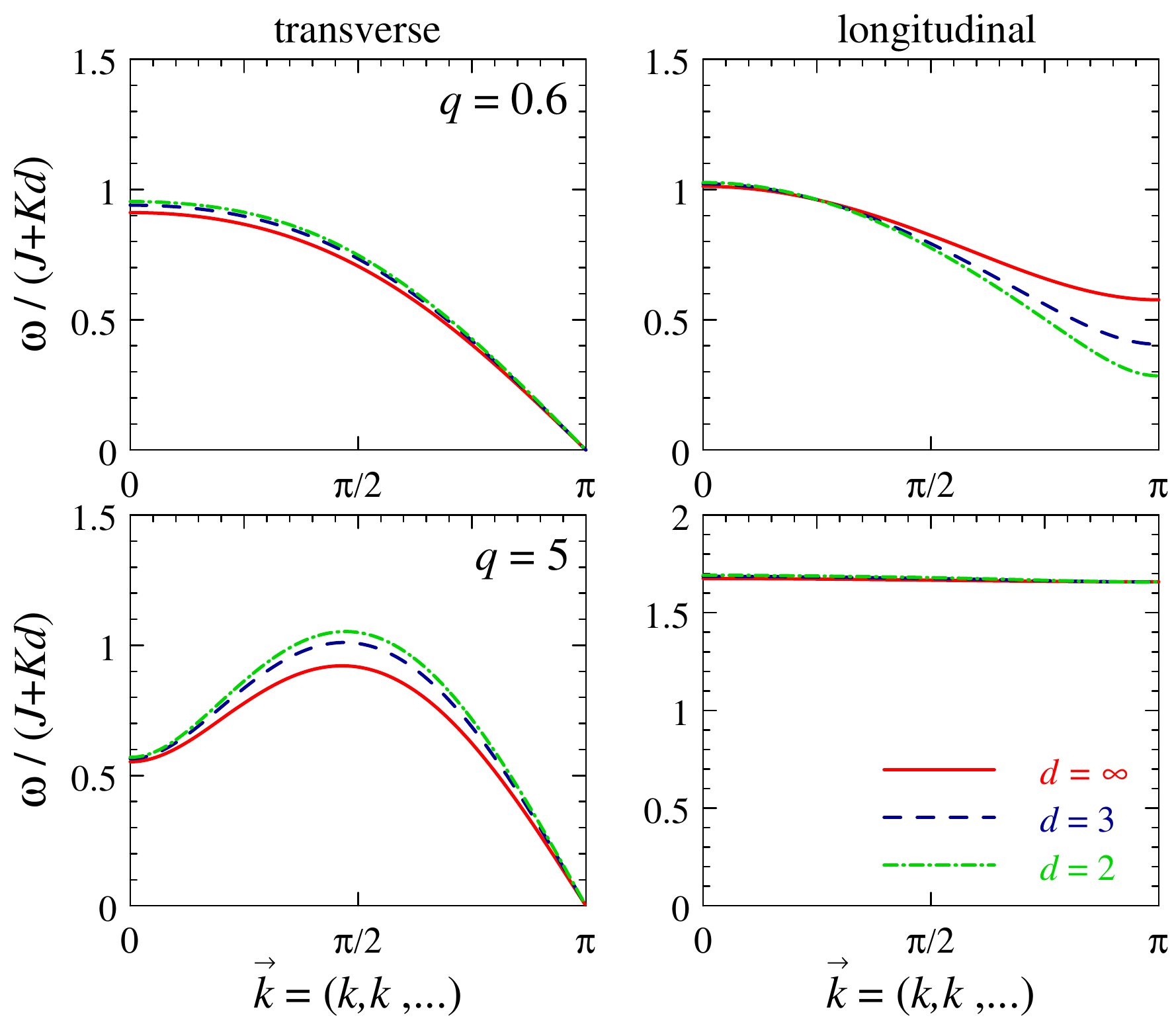}
\caption{
Dispersion of poles in the odd-channel dynamic susceptibility for $d=\infty$ (solid), $d=3$ (dashed), and $d=2$ (dash-dot), for $\qq=0.6$ (top) and $\qq=5$ (bottom).
Left: $\chi_a(\kk,\w)$ ($a=x,y$) with poles given by the transverse modes \eqref{dispa1}.
Right: $\chi_z(\kk,\w)$ with poles from the longitudinal mode \eqref{dispz1}.
The pole weights are in Eqs.~\eqref{polea} and \eqref{polez}.
Note that the energies are plotted as $\w/(J+Kd)$.
}
\label{fig:disp}
\end{figure}

As shown in I, higher orders in the $1/d$ expansion for $\chi(\kk,\w)$ place the poles at the renormalized mode frequencies $\Wt_{\vec k \alpha}$ and produce $1/d$ corrections to the weights.
Rewriting the mode weights via $\mathcal{Z}^{(o)}_{\vec k \alpha} = (J/\Wt_{\vec k \alpha}) \mathcal{W}^{(o)}_{\vec k \alpha}$ and
$\mathcal{Z}^{(e)}_{\vec k+\vec Q \alpha} = (J/\Wt_{\vec k \alpha}) \mathcal{W}^{(e)}_{\vec k+\vec Q \alpha}$, we finally find for $\hz=0$
\begin{align}
\label{polea}
\mathcal{W}^{(o)}_{\vec k a} &= \frac{(2\qq+1)[2q(1-\gk)+1+\gk]}{8\qq} + \mathcal{O}\left(\frac{1}{d}\right), \\
\mathcal{W}^{(o)}_{\vec k z} &= \frac{1}{2\qq} + \mathcal{O}\left(\frac{1}{d}\right)
\label{polez}
\end{align}
in the odd channel, and
\begin{align}
\label{poleeven}
\mathcal{W}^{(e)}_{\vec k + \vec Q a} &= \frac{(2\qq-1)(2\qq+1)(1+\gk)}{8q} + \mathcal{O}\left(\frac{1}{d}\right),
\end{align}
in the even channel where the momentum shift by $\vec{Q}$ has been made explicit.
While the transverse modes, which are gapless at the ordering wavevector $\vec{Q}$, show up in the odd channel with a spectral weight diverging $\propto 1/\w$, their $\vec{Q}$-shifted replica is seen in the even channel, but here the weight vanishes $\propto \w$ due to the factor $(1+\gk)$ in Eq.~\eqref{poleeven}.

The pole dispersion for both the transverse and longitudinal susceptibility in the odd channel is illustrated in Fig.~\ref{fig:disp}. This appears qualitatively consistent with results from series expansions for the bilayer Heisenberg model,\cite{hamer12} noting the $\vec Q$ momentum shift in the even channel.
Two things concerning the mode dispersion deep in the ordered phase are worth noticing: the longitudinal mode has a very weak dispersion, and the transverse modes develop a second dispersion minimum at $\kk = 0$. The corresponding extreme limit of vanishing intra-dimer coupling, $\qq\to\infty$, will be discussed in the next section.


\section{Vanishing intra-dimer coupling and spin-wave theory}
\label{sec:spinw}

For $J=0$ the system described by Eq.~\eqref{hh} consists of two decoupled hypercubic antiferromagnets. In this limit, conventional spin-wave theory provides a natural approach, and we will show that our $1/d$ expansion produces results consistent with those of spin waves. We will also comment on the role of the longitudinal mode in the $J=0$ limit.

\subsection{Spin waves and $1/d$ expansion}

Spin-wave theory represents an expansion around a perfectly ordered state of a spin-$S$ quantum magnet, with the small parameter being $1/S$. A convenient formulation utilizes the Holstein-Primakoff representation\cite{hp1940} of spin operators which is used to generate a Hamiltonian of interacting bosons.
Here we apply spin-wave theory to a spin-$S$ Heisenberg model on a hypercubic lattice in $d$ dimensions with nearest-neighbor interaction $K$ and demonstrate that it can be used to generate a $1/d$
expansion. As the formalism is standard\cite{yosida_book} we only quote the results.

We start with the ground-state energy per spin. To order $1/S$ the spin-wave expression reads\cite{yosida_book,hamer}
\begin{equation}
\frac{\EE^{\rm SW}}{N} = -\frac{KzS^{2}}{2} \left[ 1 + \frac{1}{S} \left( 1 - \frac{2}{N} \sum'_{\kk} \sqrt{1- \gk^{2}} \right) \right] \,,
\end{equation}
where $\gk$ is defined in Eq.~\eqref{gammadef}, $z=2d$ the coordination number, $N$ the number of sites, and the momentum summation is now over the reduced Brillouin zone of the antiferromagnetic state.
A $1/d$ expansion of this result can be generated by expanding the argument of the momentum sum in powers of $\gk$, see I for an extensive discussion. Using $(2/N) \sum'_{\kk} \gk^{2} = 1/(2d)$
we eventually find
\begin{equation}
\label{esw}
\frac{\EE^{\rm SW}}{N} = - KdS^{2} \left[ 1 + \frac{1}{S} \left( \frac{1}{4d} + \mathcal{O}\left(\frac{1}{d^2}\right) \right) \right] \,.
\end{equation}
In a similar way, we can obtain an expansion for the staggered magnetization per spin. The general $\mathcal{O}(1/S)$  expression reads:
\begin{equation}
\MM^{\rm SW} = S \left[ 1 - \frac{1}{2S} \left( \frac{2}{N} \sum'_{\kk} \frac{1}{\sqrt{1-\gk^{2}}} -1 \right) \right] \,.
\end{equation}
Expanding in powers of $\gk$ under the momentum sum yields:
\begin{equation}
\label{msw}
\MM^{\rm SW} = S \left[ 1 - \frac{1}{S} \left( \frac{1}{8d} + \mathcal{O}\left(\frac{1}{d^2}\right) \right)\right] \,.
\end{equation}
For both $\EE$ and $\MM$ it can be shown that higher-order terms in the $1/S$ expansion\cite{hamer} are suppressed at least as $1/d^2$ in the large-$d$ limit. This implies that spin-wave theory to order $1/S$ is sufficient to generate the first two terms of the $1/d$ expansion for {\em arbitrary} value of $S$.

\subsection{Bond-operator theory for vanishing intra-dimer coupling}

We can compare the above expressions with the results from the $1/d$ expansion. The ground-state energy per dimer, Eq.~\eqref{e0f}, in the limit $\qq\to\infty$ reduces to $\EE/(\qq J N) = -1/2 - 1/(4d)$, while the staggered magnetization per dimer, Eq.~\eqref{msq}, becomes in the same limit $\MM=1-1/(4d)$. Considering that a dimer consists of two spins, these results match the spin-wave results in Eqs.~\eqref{esw} and \eqref{msw} if evaluated for $S=1/2$.

In addition, one may compare the leading-order (transverse) mode dispersions in both approaches which again yields perfect agreement. In particular, the two transverse modes of bond-operator approach obey $\wt_{\kk}=\wt_{\kk+\vec{Q}}$ and are gapless both at $\kk=0$ and $\kk=\vec{Q}$. This yields a total of four Goldstone modes, which is the correct number for two independent subsystems which are collinearly ordered. (At any finite $J$ there are only two Goldstone modes.) Moreover, the mode weights in the even and odd channel are identical in the $\qq\to\infty$ limit: This is required because the fluctuations in the two layers are independent.

The comparison so far suggests that the present bond-operator theory on the one hand and spin-wave theory on the other hand are identical in the limit of decoupled ``layers'', at least if applied for large $d$. This, however, is a somewhat superficial conclusion: In bond-operator theory there is a longitudinal mode which has no counterpart in the spin-wave approach. According to its definition, this mode involves simultaneous excitations in both layers, which appears unnatural for $J=0$. Consequently, the longitudinal mode is dispersionless in this limit, $\Wt_{\kk z}=2Kd$ \eqref{dispz1}, and carries zero spectral weight, Eq.~\eqref{sozk}.

This does, however, {\em not} imply that the longitudinal mode can be discarded when performing bond-operator calculations for $J=0$. The self-energy of the transverse modes also involves longitudinal-mode propagators, see Appendix~\ref{app:selfe}. These self-energy contributions are non-vanishing and are required to fulfill the Goldstone condition, $\Wt_{\vec{Q} a}=0$, at order $1/d$. This also implies that higher-order calculations in the two approaches generically involve different intermediate quantities, diagrams etc., whereas final results are expected to match.


\section{Summary}
\label{sec:concl}

We have demonstrated that the large-$d$ expansion for coupled-dimer magnets, introduced in I, can be applied to magnetically ordered phases. It delivers consistent zero-temperature results,
order by order in $1/d$, across the entire phase diagram including the quantum critical point and its vicinity. Explicit results have been given for coupled dimers on a hypercubic lattice.
We have also shown that, in the extreme limit of vanishing intra-dimer coupling where longitudinal fluctuations do not enter most observables, our leading-order $1/d$ corrections agree
with those derived from non-linear spin-wave theory.

The success of our method shows that the bond-operator formalism, originally developed as efficient but uncontrolled mean-field theory,\cite{bondop} can be cast into a controlled and systematic theory for coupled-dimer magnets. Applications to field-induced quantum phase transitions and to systems with geometric frustration are foreseen.

On the methodological side, an interesting direction is to generalize the systematic approach presented here to finite temperatures. Inside the ordered phase,
the challenge lies in finding a suitable {\em temperature-dependent} reference state, with the condensate vanishing as the N\'{e}el temperature is approached from below.


\acknowledgments

We thank E. Andrade, S. Burdin, F. H. L. Essler, D. K. Morr, G. S. Uhrig, and M. E. Zhitomirsky for helpful discussions
as well as K. Coester and K. P. Schmidt for collaborations on related work.
This research has been supported by the DFG (GRK 1621 and SFB 1143), the GIF (G 1025-36.14/2009), and by the Virtual Institute VI-521 of the Helmholtz association.


\appendix

\section{Momentum sums and expectation values}
\label{app:sum}

Here we quote momentum sums over combinations of Bogoliubov coefficients \eqref{eq3}, as used in the main text, to $\mathcal{O}(1/d)$. According to our philosophy of $1/d$ expansion, we shall then expand these coefficients in powers of $\gk$ inside the summation to extract a $1/d$ expansion.
For instance, expansion in $\gk$ gives
\begin{align}
\ut_{\kk a}\vt_{\kk a} &= -\frac{J\qq \gk}{2\A} + \frac{J^2 \qq^2 \gk^2}{2 \A^2}\frac{1-\lm_{0}^2}{1+\lm_{0}^2} + \mathcal{O}(\gk^3)\,, \\
\ut_{\kk z}\vt_{\kk z} &= -\frac{J\qq \gk}{2\Ap}\left(\frac{1-\lm_{0}^2}{1+\lm_{0}^2}\right)^2
+ \frac{J^2 \qq^2 \gk^2}{2\Ap^2}\left(\frac{1-\lm_{0}^2}{1+\lm_{0}^2}\right)^4 + \mathcal{O}(\gk^3) \,.
\end{align}
with $\A$ and $\Ap$ defined in Eq.~\eqref{a}.
Now using the properties of momentum sums of $\gk$ i.e.
\begin{equation}
\frac{1}{N}\sum_{\kk} \gk = 0 \,,~~~~~ \frac{1}{N}\sum_{\kk} \gk^2 = \frac{1}{2d} \,,
\end{equation}
etc. we get
\begin{align}
\frac{1}{N} \sum_{\kk} \ut_{\kk a} \vt_{\kk a} &= \frac{J^{2} \qq^2}{4 \A^{2} d}\frac{1-\lm_{0}^2}{1+\lm_{0}^2}  + \mathcal{O}(d^{-2}) \,,
\\
\frac{1}{N} \sum_{\kk} \ut_{\kk z} \vt_{\kk z} &= \frac{J^{2} \qq^2}{4 \Ap^{2} d}\left(\frac{1-\lm_{0}^2}{1+\lm_{0}^2}\right)^4   + \mathcal{O}(d^{-2}) \,.
\end{align}
Similarly other combinations of Bogoliubov coefficients when summed over $\kk$ give a $1/d$ expansion. Following is the summary of relevant momentum sums:
\begin{align}
\label{r1}
R_{1a} &= \frac{1}{N} \sum_{\kk} \ut_{\kk a} \vt_{\kk a} = \frac{J^{2} \qq^2}{4 \A^{2} d}\frac{1-\lm_{0}^2}{1+\lm_{0}^2}  + \mathcal{O}(d^{-2}) \\ 
\label{r2}
R_{2a} &= \frac{1}{N} \sum_{\kk} \vt^{2}_{\kk a} = \frac{\qq^2 J^{2}}{8 \A^{2} d}  + \mathcal{O}(d^{-2}) \\ 
\label{r3}
R_{3a} &= \frac{1}{N} \sum_{\kk} \gk \ut_{\kk a} \vt_{\kk a} = -\frac{\qq J}{4 \A d} + \mathcal{O}(d^{-2}) \\ 
\label{r4}
R_{4a} &= \frac{1}{N} \sum_{\kk} \gk \vt^{2}_{\kk m} = \mathcal{O}(d^{-2})
\end{align}
\begin{align}
\label{r1z}
R_{1z} &= \frac{1}{N} \sum_{\kk} \ut_{\kk z} \vt_{\kk z} = \frac{J^{2} \qq^2}{4 \Ap^{2} d}\left(\frac{1-\lm_{0}^2}{1+\lm_{0}^2}\right)^4  + \mathcal{O}(d^{-2}) \\ 
\label{r2z}
R_{2z} &= \frac{1}{N} \sum_{\kk} \vt^{2}_{\kk z} = \frac{J^{2} \qq^2}{8 \Ap^{2} d}\left(\frac{1-\lm_{0}^2}{1+\lm_{0}^2}\right)^4  + \mathcal{O}(d^{-2}) \\ 
\label{r3z}
R_{3z} &= \frac{1}{N} \sum_{\kk} \gk \ut_{\kk z} \vt_{\kk z} = -\frac{J \qq}{4 \Ap d}\left(\frac{1-\lm_{0}^2}{1+\lm_{0}^2}\right)^2 + \mathcal{O}(d^{-2}) \\ 
\label{r4z}
R_{4z} &= \frac{1}{N} \sum_{\kk} \gk \vt^{2}_{\kk z} = \mathcal{O}(d^{-2})
\end{align}
Note that these expressions are valid for arbitrary $\hz$, with its value entering via $\lm_0(\hz)$ according to Eq.~\eqref{h1a}.

The $R_{1\ldots4}$ are related to expectation values of the leading-order bilinear Hamiltonian \eqref{H2k} as
follows:
\begin{align}
\sum_{i} \langle t_{i\alpha}^\dagger t_{i\beta}^\dagger \rangle &= 3N \delta_{\alpha\beta} R_{1 \alpha} \,,~
\sum_{i} \langle t_{i\alpha}^\dagger t_{i\beta} \rangle = 3N \delta_{\alpha\beta} R_{2 \alpha}, \notag \\
\sum_{\langle ij \rangle} \langle t_{i\alpha}^\dagger t_{j\beta}^\dagger \rangle &= 3Nd \delta_{\alpha\beta} R_{3 \alpha} \,,
\sum_{\langle ij \rangle} \langle t_{i\alpha}^\dagger t_{j\beta} \rangle = 3Nd \delta_{\alpha\beta} R_{4 \alpha} \,.
\end{align}
Within self-energy expressions we also need
\begin{equation}
\label{raz}
R'_{az}(\kk) = \frac{1}{N}\sum_{\kk'} \ut_{\kk' a} \vt_{\kk' a} \ut_{(\kk'-\kk)z}\vt_{(\kk'-\kk)z} = \frac{J^{2} \gk}{32 \A \Ap d} \,.
\end{equation}

Similar to I, the anomalous expectation value $\langle \tilde{t}_{i\alpha}^\dagger \tilde{t}_{i\alpha}^\dagger \rangle$, being finite at the harmonic level, vanishes upon taking into account $1/d$ corrections as required by the constraint.\cite{corrfoot}


\begin{widetext}

\section{Hamiltonian coefficients}
\label{app:vert}

We start by listing the coefficients of $\mathcal{H}'_{2c}$, representing the bilinear terms arising from normal ordering of quartic interactions.

\begin{align}
\label{eq:cka}
C_{\kk a} &= (\ut_{\kk a}^{2} + \vt_{\kk a}^{2}) \qq J \left[ -2\gk R_{1a} - 6 R_{3a} - 6(\gk R_{2a} + R_{4a})\Lm \right.
- 4(R'_{4a}+2 R_{2a})\frac{\lm^{2}}{(1+\lm^{2})^{2}} + R'_{4a}  \nonumber \\
&+(R'_{4z}- 2\gk  R_{2z}) \Lm       
\left. - 2 (R_{3z}+R_{4z})\Lm^{2} - 8 R_{2z} \frac{\lm^{2}}{(1+\lm^{2})^{2}} \right] \nonumber \\
&-2\ut_{\kk a}\vt_{\kk a} \qq J \left[ 6\gk R_{2a}+ 2R_{4a} + 2(\gk R_{1a}+R_{3a})\Lm + 4R'_{3a}\frac{\lm^{2}}{(1+\lm^{2})^{2}}
+R'_{3a} + 2\gk R_{2z} + R'_{3z} \right] \,, \\
\label{dka}
D_{\kk a} &= -(\ut_{\kk a}^{2} + \vt_{\kk a}^{2})\qq J \left[ 6\gk R_{2a}+ 2R_{4a} + 2(\gk R_{1a}+R_{3a})\Lm + 4R'_{3a}\frac{\lm^{2}}{(1+\lm^{2})^{2}}
+R'_{3a} + 2\gk R_{2z} + R'_{3z} \right] \nonumber \\
&+2\ut_{\kk a}\vt_{\kk a} \qq J \left[ -2\gk R_{1a} - 6 R_{3a} - 6(\gk R_{2a} + R_{4a})\Lm \right.
- 4(R'_{4a}+2 R_{2a})\frac{\lm^{2}}{(1+\lm^{2})^{2}} + R'_{4a}  \nonumber \\
&+ (R'_{4z}- 2\gk R_{2z}) \Lm
\left. - 2 (R_{3z}+R_{4z})\Lm^{2} - 8 R_{2z} \frac{\lm^{2}}{(1+\lm^{2})^{2}} \right]\,, \\
\label{ckz}
C_{\kk z} &= -(\ut_{\kk z}^{2} + \vt_{\kk z}^{2})\qq J \left[  (2\gk R_{1z} +4R_{3z}+4R_{4z}+4\gk R_{2z})\Lm^{2}  \right.
+ 16(R_{2z}+R'_{4z})\frac{\lm^{2}}{(1+\lm^{2})^{2}} +4R_{3a} \nonumber \\
&\left.+ 2(2R_{4a}-R'_{4a})\Lm + 4\gk R_{2a} \Lm^{2} + 16R_{2a} \frac{\lm^{2}}{(1+\lm^{2})^{2}} \right] \nonumber \\
&-2\ut_{\kk z}\vt_{\kk z} \qq J \left[ (4\gk R_{2z} + 2 R_{4z} + 2R_{3z} + 2\gk R_{1z})v^{2} + 16 R'_{3z} \frac{\lm^{2}}{(1+\lm^{2})^{2}}
+2R'_{3a} + 4\gk R_{2a}\Lm^{2}\right] \,, \\
\label{dkz}
D_{\kk z} &= -(\ut_{\kk z}^{2} + \vt_{\kk z}^{2})\qq J \left[ (4\gk R_{2z} + 2 R_{4z} + 2R_{3z} + 2\gk R_{1z})\Lm^{2}
+ 16 R'_{3z} \frac{\lm^{2}}{(1+\lm^{2})^{2}}
+2R'_{3a} + 4\gk R_{2a}\Lm^{2}\right] \nonumber \\
&-2\ut_{\kk z}\vt_{\kk z} \qq J\left[ (2\gk R_{1z} +4R_{3z}+4R_{4z}+4\gk R_{2z}) \Lm^{2} \right.
+ 16(R_{2z}+R'_{4z})\frac{\lm^{2}}{(1+\lm^{2})^{2}} +4R_{3a} \nonumber \\
&\left. + 2(2R_{4a}-R'_{4a})\Lm + 4\gk R_{2a} \Lm^{2} + 16 R_{2a} \frac{\lm^{2}}{(1+\lm^{2})^{2}} \right] \,,
\end{align}
where $R$'s are momentum summations of some combination of Bogoliubov coefficients (see Appendix \ref{app:sum}) and $\Lm = \frac{1-\lm^{2}}{1+\lm^{2}}$.
The cubic vertices entering $\mathcal{H}'_3$ read:
\begin{align}
\label{vcxa}
\vcxa &= (\Ja \gamma_{2+3} +\Jb \gamma_{1+3} + h_{1a}(\lm,\hz))(\ut_{1a} \ut_{2z} \vt_{3a} + \vt_{1a} \vt_{2z} \ut_{3a}) \,, \\
\label{vcxb}
\vcxb &= (\Ja \gamma_{3-2} +\Jb \gamma_{1+3} + h_{1a}(\lm,\hz))(\ut_{1a} \vt_{2z} \vt_{3a} + \vt_{1a} \ut_{2z} \ut_{3a}) \,, \\
\label{vcxc}
\vcxc &= (\Ja \gamma_{2-3} +\Jb \gamma_{1-3} + h_{1a}(\lm,\hz))(\ut_{1a} \ut_{2z} \ut_{3a} + \vt_{1a} \vt_{2z} \vt_{3a}) \nonumber \\
&+ (\Ja \gamma_{1+2} +\Jb \gamma_{1-3} + h_{1a}(\lm,\hz))(\vt_{1a} \ut_{2z} \vt_{3a} + \ut_{1a} \vt_{2z} \ut_{3a}) \,, \\
\label{vcza}
\vcza &= (2 \Jb \gamma_{2+3} + h_{1a}(\lm,\hz))(\ut_{1z} \ut_{2z} \vt_{3z} + \vt_{1z} \vt_{2z} \ut_{3z}), \,, \\
\label{vczb}
\vczb &= (2\Jb \gamma_{2-3} + h_{1a}(\lm,\hz))(\ut_{1z} \ut_{2z} \ut_{3z} + \ut_{1z} \vt_{2z} \vt_{3z} + \vt_{1z} \ut_{2z} \ut_{3z} + \vt_{1z} \vt_{2z} \vt_{3z}) \nonumber \\
&+ (2\Jb \gamma_{1+2} + h_{1a}(\lm,\hz))(\vt_{1z} \ut_{2z} \vt_{3z} + \ut_{1z} \vt_{2z} \ut_{3z}) \,.
\end{align}

The expressions for the quartic vertices are lengthy, and in the following we only show selected ones:
\begin{align}
\label{vqaza}
\vqaza &= -\qq J \left[ \gamma_{2+3+4}(\ut_{1a}\ut_{2a}\ut_{3z}\vt_{4z} + \vt_{1a}\vt_{2a}\vt_{3z}\ut_{4z}) + \Lm \gamma_{2+3+4} \right.
(\ut_{1a}\vt_{2a}\vt_{3z}\ut_{4z} + \vt_{1a}\ut_{2a}\ut_{3z}\vt_{4z}) \nonumber \\
&+\frac{\gamma_{2+4}}{2} (\ut_{1a}\ut_{2a}\vt_{3z}\vt_{4z} + \vt_{1a}\vt_{2a}\ut_{3z}\ut_{4z})
- \Lm \frac{\gamma_{2+4}}{2} (\ut_{1a}\vt_{2a}\vt_{3z}\ut_{4z} + \vt_{1a}\ut_{2a}\ut_{3z}\vt_{4z}) \nonumber \\
&+\Lm^{2} (\gamma_{3} \ut_{1a}\vt_{2a}\ut_{3z}\ut_{4z} + \gamma_{3} \vt_{1a}\ut_{2a}\vt_{3z}\vt_{4z}
+\gamma_{4} \vt_{1a}\ut_{2a}\ut_{3z}\vt_{4z} + \gamma_{4} \ut_{1a}\vt_{2a}\vt_{3z}\ut_{4z}) \nonumber \\
&\left. +\frac{4\lm^{2} \gamma_{1+2}}{(1+\lm^{2})^{2}}  (\vt_{1a}\ut_{2a}\vt_{3z}\ut_{4z} + \ut_{1a}\vt_{2a}\ut_{3z}\vt_{4z}) \right],
\end{align}
\begin{align}
\label{vqaze}
\vqaze &= -\qq J \left[ \Lm \gamma_{2+3-4}(\ut_{1a}\vt_{2a}\ut_{3z}\ut_{4z}
+ \ut_{1a}\vt_{2a}\vt_{3z}\vt_{4z}+\vt_{1a}\ut_{2a}\ut_{3z}\ut_{4z} + \vt_{1a}\ut_{2a}\vt_{3z}\vt_{4z}) \right. \nonumber \\
&+\gamma_{2+3-4}(\ut_{1a}\ut_{2a}\ut_{3z}\ut_{4z} + \ut_{1a}\ut_{2a}\vt_{3z}\vt_{4z} + \vt_{1a}\vt_{2a}\ut_{3z}\ut_{4z} + \vt_{1a}\vt_{2a}\vt_{3z}\vt_{4z}) \nonumber \\
&+\frac{\gamma_{2-4}}{2} (\ut_{1a}\ut_{2a}\vt_{3z}\ut_{4z} + \vt_{1a}\vt_{2a}\ut_{3z}\vt_{4z})
+\frac{\gamma_{2+3}}{2} (\ut_{1a}\ut_{2a}\vt_{3z}\ut_{4z} + \vt_{1a}\vt_{2a}\ut_{3z}\vt_{4z}) \nonumber \\
&- \frac{\Lm}{2}(\gamma_{2+3}\ut_{1a}\vt_{2a}\ut_{3z}\ut_{4z}+\gamma_{2+3}\vt_{1a}\ut_{2a}\vt_{3z}\vt_{4z}
+ \gamma_{2-4}\ut_{1a}\vt_{2a}\vt_{3z}\vt_{4z} +  \gamma_{2-4}\vt_{1a}\ut_{2a}\ut_{3z}\ut_{4z}) \nonumber \\
&+\Lm^{2} (\gamma_{3} \ut_{1a}\vt_{2a}\ut_{3z}\vt_{4z} + \gamma_{3} \vt_{1a}\ut_{2a}\vt_{3z}\ut_{4z}
+\gamma_{3} \vt_{1a}\ut_{2a}\vt_{3z}\vt_{4z} + \gamma_{3} \ut_{1a}\vt_{2a}\ut_{3z}\ut_{4z} \nonumber \\
&~~~~~~~~~~~~~~~~~+\gamma_{4} \ut_{1a}\vt_{2a}\ut_{3z}\vt_{4z} + \gamma_{4} \vt_{1a}\ut_{2a}\vt_{3z}\ut_{4z}
+\gamma_{4} \vt_{1a}\ut_{2a}\ut_{3z}\ut_{4z} + \gamma_{4} \ut_{1a}\vt_{2a}\vt_{3z}\vt_{4z}) \nonumber \\
&\left. +\frac{4\lm^{2}}{(1+\lm^{2})^{2}} \gamma_{1+2} (\vt_{1a}\ut_{2a}\ut_{3z}\ut_{4z} + \vt_{1a}\ut_{2a}\vt_{3z}\vt_{4z}
+\ut_{1a}\vt_{2a}\ut_{3z}\ut_{4z} + \ut_{1a}\vt_{2a}\vt_{3z}\vt_{4z}) \right], \\
\label{vqazf}
\vqazf &= -\qq J \left[ \Lm (\gamma_{1+2-4} \vt_{1z}\ut_{2z}\ut_{3a}\ut_{4a}+ \gamma_{1+2-4} \ut_{1z}\vt_{2z}\vt_{3a}\vt_{4a}
+\gamma_{1+2+3}\vt_{1z}\ut_{2z}\vt_{3a}\vt_{4a} + \gamma_{1+2+3} \ut_{1z}\vt_{2z}\ut_{3a}\ut_{4a}) \right. \nonumber \\
&+(\gamma_{1+2-4} \ut_{1z}\vt_{2z}\ut_{3a}\vt_{4a}+ \gamma_{1+2-4} \vt_{1z}\ut_{2z}\vt_{3a}\ut_{4a}
+\gamma_{1+2+3}\ut_{1z}\vt_{2z}\ut_{3a}\vt_{4a} + \gamma_{1+2+3} \vt_{1z}\ut_{2z}\vt_{3a}\ut_{4a}) \nonumber \\
&+\frac{\gamma_{2-4}}{2} (\vt_{1z}\vt_{2z}\ut_{3a}\vt_{4a} + \ut_{1z}\ut_{2z}\vt_{3a}\ut_{4a})
+\frac{\gamma_{2+3}}{2} (\vt_{1z}\vt_{2z}\ut_{3a}\vt_{4a} + \ut_{1z}\ut_{2z}\vt_{3a}\ut_{4a}) \nonumber \\
&- \frac{\Lm}{2}(\gamma_{2+3}\vt_{1z}\ut_{2z}\vt_{3a}\vt_{4a}+\gamma_{2+3}\ut_{1z}\vt_{2z}\ut_{3a}\ut_{4a}
+ \gamma_{2-4}\vt_{1z}\ut_{2z}\ut_{3a}\ut_{4a}  + \gamma_{2-4}\ut_{1z}\vt_{2z}\vt_{3a}\vt_{4a}) \nonumber \\
&+\Lm^{2} (\gamma_{1} \ut_{1z}\ut_{2z}\ut_{3a}\ut_{4a} + \gamma_{1}\ut_{1z}\ut_{2z}\vt_{3a}\vt_{4a}
+\gamma_{1}\vt_{1z}\vt_{2z}\ut_{3a}\ut_{4a} + \gamma_{1}\vt_{1z}\vt_{2z}\vt_{3a}\vt_{4a} \nonumber \\
&~~~~~~~~~~~~~~~~~+\gamma_{2} \ut_{1z}\vt_{2z}\ut_{3a}\ut_{4a} + \gamma_{2} \ut_{1z}\vt_{2z}\vt_{3a}\vt_{4a}
+\gamma_{2} \vt_{1z}\ut_{2z}\ut_{3a}\ut_{4a} + \gamma_{2} \vt_{1z}\ut_{2z}\vt_{3a}\vt_{4a}) \nonumber \\
&\left. +\frac{4\lm^{2}}{(1+\lm^{2})^{2}} \gamma_{3-4} (\ut_{1z}\vt_{2z}\ut_{3a}\ut_{4a} + \ut_{1z}\vt_{2z}\vt_{3a}\vt_{4a}
+\vt_{1z}\ut_{2z}\ut_{3a}\ut_{4a} + \vt_{1z}\ut_{2z}\vt_{3a}\vt_{4a}) \right], \\
\label{vqza}
\vqza &= -\qq J \Lm^{2} \gamma_{1} (\ut_{1z}\ut_{2z}\ut_{3z}\vt_{4z}+\vt_{1z}\vt_{2z}\vt_{3z}\ut_{4z})
-\qq J \left(\gamma_{4} \Lm^{2} + \frac{4\gamma_{2+4} \lm^{2}}{(1+\lm^{2})^{2}}\right) (\ut_{1z}\ut_{2z}\vt_{3z}\vt_{4z}+\vt_{1z}\vt_{2z}\ut_{3z}\ut_{4z}), \\
\label{vqzc}
\vqzc &= -\qq J \Lm^{2} \left[ \gamma_{1} (\ut_{1z}\ut_{2z}\ut_{3z}\ut_{4z}+\ut_{1z}\ut_{2z}\vt_{3z}\vt_{4z}
+\ut_{1z}\vt_{2z}\ut_{3z}\vt_{4z}+\vt_{1z}\ut_{2z}\vt_{3z}\ut_{4z}  \right. \nonumber \\
&~~~~~~~~~~~~~~~~~~~~~~~~+ \vt_{1z}\vt_{2z}\ut_{3z}\ut_{4z}+\vt_{1z}\vt_{2z}\vt_{3z}\vt_{4z})
\left.+\gamma_{4} (\vt_{1z}\ut_{2z}\ut_{3z}\vt_{4z}+\ut_{1z}\vt_{2z}\vt_{3z}\ut_{4z}) \right] \nonumber \\
&-\frac{\qq J}{(1+\lm^{2})^{2}} \left[ \left(\gamma_{4} (1-\lm^{2})^{2} + 4\gamma_{2-4} \lm^{2}\right) (\ut_{1z}\ut_{2z}\vt_{3z}\ut_{4z}+\vt_{1z}\vt_{2z}\ut_{3z}\vt_{4z}) \right. \nonumber \\
&~~~~~~~~~~~~~~~+\left(\gamma_{3} (1-\lm^{2})^{2} + 4\gamma_{2+3} \lm^{2}\right) (\ut_{1z}\ut_{2z}\vt_{3z}\ut_{4z}+\vt_{1z}\vt_{2z}\ut_{3z}\vt_{4z}) \nonumber \\
&~~~~~~~~~~~~~~~+\left(\gamma_{2} (1-\lm^{2})^{2} + 4\gamma_{2-4} \lm^{2}\right) (\ut_{1z}\vt_{2z}\vt_{3z}\vt_{4z}+\vt_{1z}\ut_{2z}\ut_{3z}\ut_{4z}) \nonumber \\
&\left.~~~~~~~~~~~~~~+\left(\gamma_{1} (1-\lm^{2})^{2} + 4\gamma_{1+2} \lm^{2}\right) (\vt_{1z}\ut_{2z}\vt_{3z}\vt_{4z}+\ut_{1z}\vt_{2z}\ut_{3z}\ut_{4z})\right], \\
\label{vqaba}
\vqaba &= -\qq J \gamma_{2+3+4} (\ut_{1a}\ut_{2a}\ut_{3b}\vt_{4b} + \vt_{1a}\vt_{2a}\vt_{3b}\ut_{4b} )
-\qq J \Lm \gamma_{2+3+4} (\ut_{1a}\vt_{2a}\vt_{3b}\ut_{4b} + \vt_{1a}\ut_{2a}\ut_{3b}\vt_{4b}) \nonumber \\
&-\frac{2\qq J \lm^{2}}{(1+\lm^{2})^{2}} \gamma_{3+4} \ut_{1a}\vt_{2a}\vt_{3b}\ut_{4b}
-\frac{\qq J}{2} \gamma_{2+4} (\ut_{1a}\ut_{2a}\vt_{3b}\vt_{4b} - \ut_{1a}\vt_{2a}\vt_{3b}\ut_{4b}) (1-\delta_{ab}), \\
\label{vqabd}
\vqabd &= -\qq J \left[ \gamma_{2+3-4} (\ut_{1a}\ut_{2a}\ut_{3b}\ut_{4b} + \ut_{1a}\ut_{2a}\vt_{3b}\vt_{4b}
+ \vt_{1a}\vt_{2a}\ut_{3b}\ut_{4b} + \vt_{1a}\vt_{2a}\vt_{3b}\vt_{4b}) \right. \nonumber \\
&+ \gamma_{1+2-4} (\ut_{1a}\vt_{2a}\ut_{3b}\vt_{4b} + \vt_{1a}\ut_{2a}\vt_{3b}\ut_{4b})
\left.+ \gamma_{1+2+3}(\ut_{1a}\vt_{2a}\ut_{3b}\vt_{4b} + \vt_{1a}\ut_{2a}\vt_{3b}\ut_{4b}) \right] \nonumber \\
&-\qq J \Lm \left[ \gamma_{2+3-4} (\ut_{1a}\vt_{2a}\ut_{3b}\ut_{4b} + \ut_{1a}\vt_{2a}\vt_{3b}\vt_{4b}
+\vt_{1a}\ut_{2a}\ut_{3b}\ut_{4b} + \vt_{1a}\ut_{2a}\vt_{3b}\vt_{4b}) \right. \nonumber \\
&+\gamma_{1+2-4} (\vt_{1a}\ut_{2a}\ut_{3b}\ut_{4b} + \ut_{1a}\vt_{2a}\vt_{3b}\vt_{4b})
\left.+\gamma_{1+2+3} (\vt_{1a}\ut_{2a}\vt_{3b}\vt_{4b} + \ut_{1a}\vt_{2a}\ut_{3b}\ut_{4b}) \right] \nonumber \\
&-\frac{2\qq J \lm^{2}}{(1+\lm^{2})^{2}} \left[\gamma_{1+2}(\vt_{1a}\ut_{2a}\ut_{3b}\ut_{4b} + \vt_{1a}\ut_{2a}\vt_{3b}\vt_{4b})
+\gamma_{3-4}(\ut_{1a}\vt_{2a}\ut_{3b}\ut_{4b} + \ut_{1a}\vt_{2a}\vt_{3b}\vt_{4b}) \right] \nonumber \\
&-\frac{\qq J}{2} \left[ \gamma_{2-4} (\ut_{1a}\ut_{2a}\vt_{3b}\ut_{4b} + \vt_{1a}\vt_{2a}\ut_{3b}\vt_{4b}
- \vt_{1a}\ut_{2a}\ut_{3b}\ut_{4b} - \ut_{1a}\vt_{2a}\vt_{3b}\vt_{4b})  \right. \nonumber \\
&+\gamma_{2+3} (\ut_{1a}\ut_{2a}\vt_{3b}\ut_{4b} + \vt_{1a}\vt_{2a}\ut_{3b}\vt_{4b}
\left.- \ut_{1a}\vt_{2a}\ut_{3b}\ut_{4b} - \vt_{1a}\ut_{2a}\vt_{3b}\vt_{4b})\right](1-\delta_{ab}).
\end{align}



\section{Self-energies}
\label{app:selfe}

This appendix is devoted to the normal self-energies of the $\tb$ particles, needed for determining the mode dispersion to order $1/d$.

The self-energy diagrams for the transverse modes are shown in Fig.~\ref{fig:selft}. Evaluating the frequency and momentum integrals, we find to order $1/d$:
\begin{align}
\Sigma^{\ref{fig:selft}(a)}(\kk,\wt) &= A^{(1)}_{\kk a} (\ut_{\kk a}^{2} + \vt_{\kk a}^{2}) + 2B^{(1)}_{\kk a}\ut_{\kk a}\vt_{\kk a} + C_{\kk a} \,, \\
\Sigma^{\ref{fig:selft}(b)}(\kk,\wt) &= \frac{1}{\wt-\A-\Ap} \left[ \ut_{\kk a}^{2} \left(\frac{\Ja^{2}+\Jb^{2}-2\Ja \Jb \gk}{2d} + 2 \Ja^{2} \gk R'_{3z}(\kk-\vec{Q}) \right.
- 2\Ja \Jb \gk R'_{3z}(\vec{Q}) + \Ja^{2} \gk^{2} R_{2z}\right)  \nonumber \\
&+ \vt_{\kk a}^{2} \Ja^{2} \gk^{2} R_{2a}
\left.+2\ut_{\kk a}\vt_{\kk a} \left(\Ja^{2} \gk R_{3a} -2\Ja \Jb \gk R'_{3a}(\kk) + \Ja^{2} \gk^{2} R'_{az}(\kk-\vec{Q})\right) \right], \\
\Sigma^{\ref{fig:selft}(c)}(\kk,\wt) &= -\frac{1}{\wt+\A+\Ap} \left[ \vt_{\kk a}^{2} \left(\frac{\Ja^{2}+\Jb^{2}-2\Ja \Jb \gk}{2d} + 2 \Ja^{2} \gk R'_{3z}(\kk-\vec{Q}) \right.
- 2\Ja \Jb \gk R'_{3z}(\vec{Q}) + \Ja^{2} \gk^{2} R_{2z}\right)  \nonumber \\
&+ \ut_{\kk a}^{2} \Ja^{2} \gk^{2} R_{2a}
\left.+2\ut_{\kk a}\vt_{\kk a} \left(\Ja^{2} \gk R_{3a} -2\Ja \Jb \gk R'_{3a}(\kk) + \Ja^{2} \gk^{2} R'_{az}(\kk-\vec{Q})\right)\right], \\
\Sigma^{\ref{fig:selft}(d)}(\kk,\wt) &= \Sigma^{\ref{fig:selft}(e)}(\kk,\wt) = -\frac{\gk \qq J^{2} R_{3a}}{2 \A} \left[ \ut_{\kk a}^{2} + \vt_{\kk a}^{2}
+ 2\ut_{\kk a}\vt_{\kk a} \Lm_{0} \right],
 \\
\Sigma^{\ref{fig:selft}(f)}(\kk,\wt) &= \frac{\qq^{2} J^{2}}{\wt-\A-2\Ap} \left[ \ut_{\kk a}^{2} \left(\frac{\Lm_{0}^{4}}{2d}  \right.
+ \gk^{2}  R_{2z} \Lm_{0}^{2} +2 \gk R_{3z} \Lm_{0}^{3}\right) 
+ \vt_{\kk a}^{2} \gk^{2}  R_{2z}
\left. + 2\ut_{\kk a}\vt_{\kk a} \left(\gk^{2}  R_{2z} \Lm_{0} + \gk  R_{3z} \Lm_{0}^{2} \right) \right],
 \\
\Sigma^{\ref{fig:selft}(g)}(\kk,\wt) &= -\frac{\qq^{2} J^{2}}{\wt+\A+2\Ap} \left[ \vt_{\kk a}^{2} \left(\frac{\Lm_{0}^{4}}{2d}  \right.
+ \gk^{2}  R_{2z} \Lm_{0}^{2} + 2\gk R_{3z} \Lm_{0}^{3}\right) 
+ \ut_{\kk a}^{2} \gk^{2}  R_{2z}
\left. + 2\ut_{\kk a}\vt_{\kk a} \left(\gk^{2}  R_{2z} \Lm_{0} + \gk  R_{3z} \Lm_{0}^{2} \right) \right],
\end{align}
\end{widetext}

\vfill

\begin{figure}[b]
\includegraphics[width=0.48\textwidth]{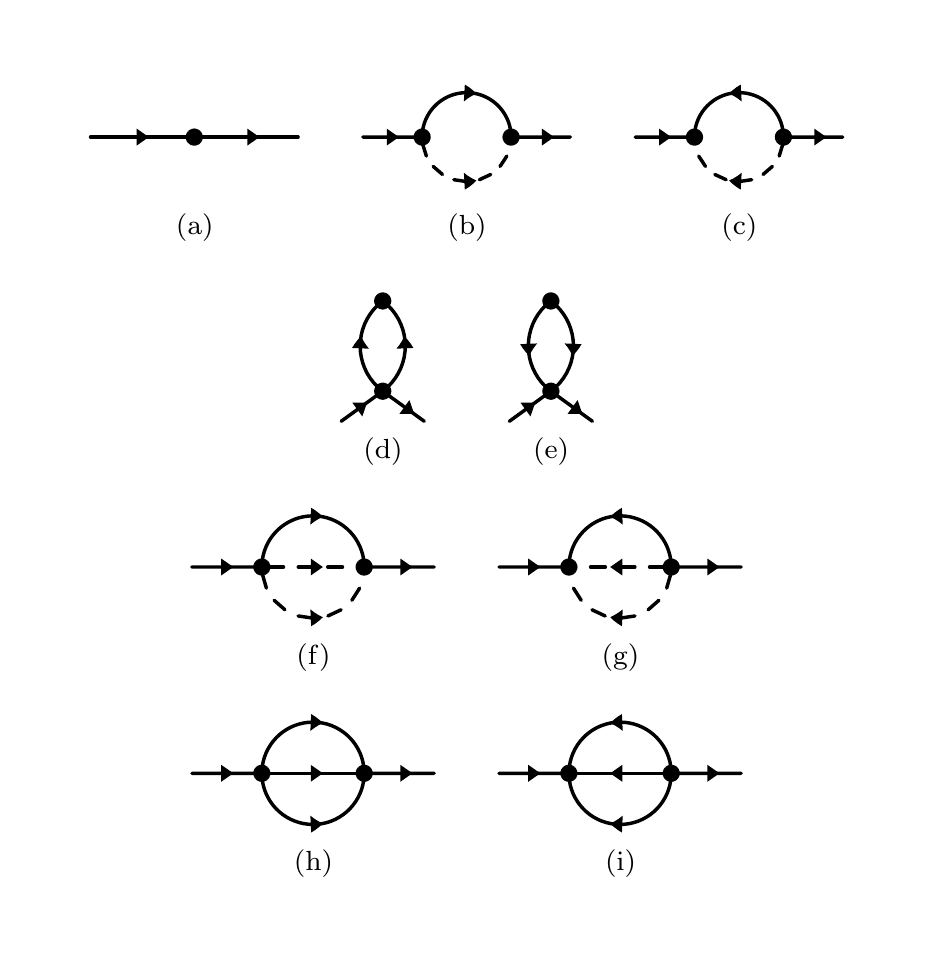}
\caption{Self-energy diagrams contributing to the transverse mode dispersion up to order $1/d$. Solid (dashed) lines correspond to $\tb_{xy}$ ($\tb_z$) propagators. The bilinear vertex represents $\mathcal{H}'_{2b}+\mathcal{H}'_{2c}$, while the cubic (quartic) vertices are for $\mathcal{H}'_{3}$ ($\mathcal{H}'_{4}$).
}
\label{fig:selft}
\end{figure}

\begin{figure}[b]
\includegraphics[width=0.48\textwidth]{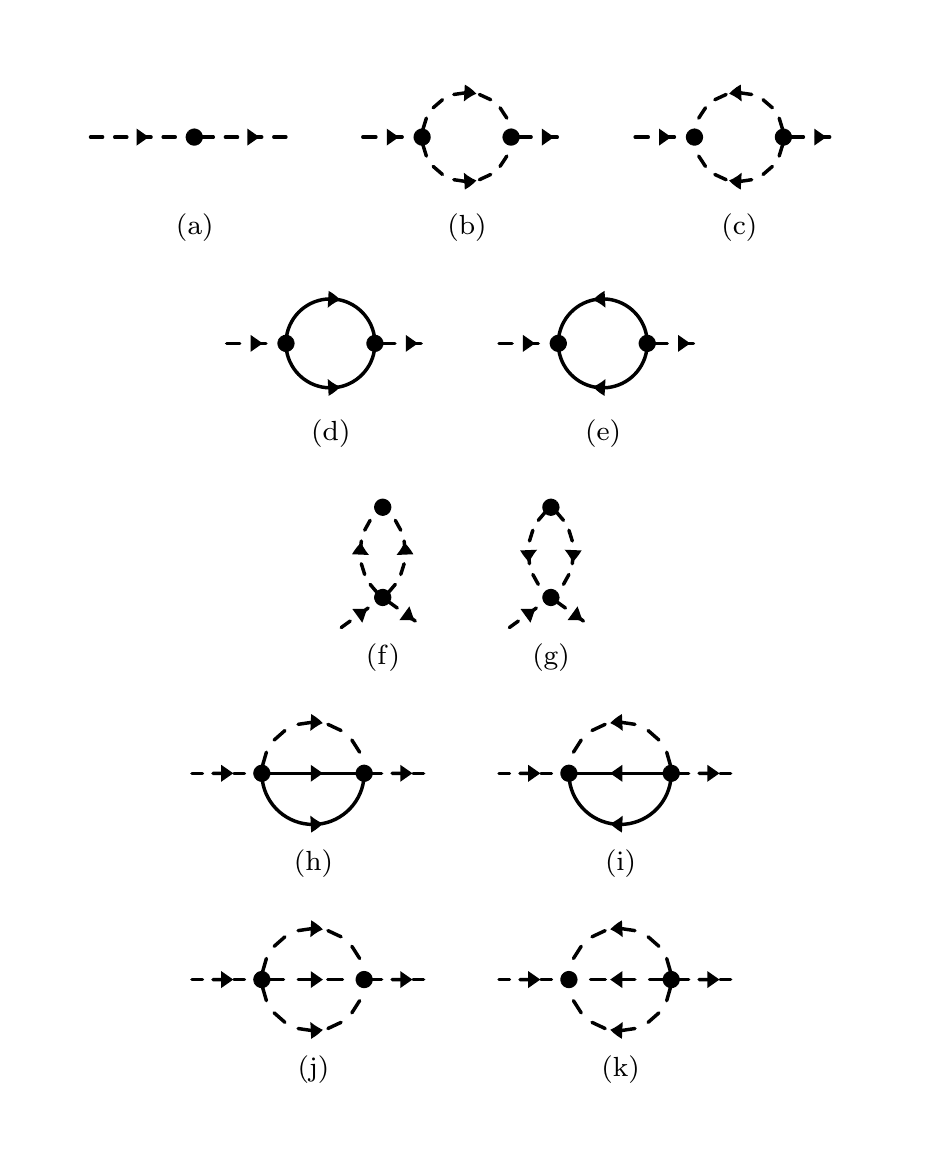}
\caption{Feynman diagrams contributing to the longitudinal mode dispersion up to order $1/d$; the notation is as in Fig.~\ref{fig:selft}.}
\label{fig:selfl}
\end{figure}

\clearpage
\begin{widetext}

\begin{align}
\Sigma^{\ref{fig:selft}(h)}(\kk,\wt) &= \frac{\qq^{2} J^{2}}{\wt-3 \A} \left[ \ut_{\kk a}^{2} \left(3\gk^{2}  R_{2a} \Lm_{0}^{2} \right.
+ 6\gk  R_{3a} \Lm_{0} + \frac{3}{2d}\right)  
+ 3\vt_{\kk a}^{2} \gk^{2}  R_{2a}
\left.+2\ut_{\kk a}\vt_{\kk a} \left(3\gk^{2}  R_{2a} \Lm_{0} + 3\gk  R_{3a}\right) \right], \\
\Sigma^{\ref{fig:selft}(i)}(\kk,\wt) &= -\frac{\qq^{2} J^{2}}{\wt+3\A} \left[ \vt_{\kk a}^{2} \left(3\gk^{2}  R_{2a} \Lm_{0}^{2} \right.
+ 6\gk  R_{3a} \Lm_{0} + \frac{3}{2d}\right)  
+ 3\ut_{\kk a}^{2} \gk^{2}  R_{2a}
\left.+2\ut_{\kk a}\vt_{\kk a} \left(3\gk^{2}  R_{2a} \Lm_{0} + 3\gk  R_{3a}\right) \right],
\end{align}
with abbreviations $\Lm_0 = \frac{1-\lm_{0}^{2}}{1+\lm_{0}^{2}}$, $\A$ and $\Ap$ from Eq.~\eqref{a} and $\Ja$ and $\Jb$ from Eq.~\eqref{j1j2}.
We recall that all Hamiltonian pieces $\mathcal{H}'_n(\lm)$ explicitly depend on the condensate parameter $\lm$; to order $1/d$ it is sufficient to evaluate the self-energy diagrams
(and thus $\Ja$ and $\Jb$) at $\lm=\lm_0$, with the only exception of the first two terms of $\Sigma^{\ref{fig:selft}(a)}$ which arise from $\mathcal{H}'_{2b}$.

The Feynman diagrams for longitudinal-mode self-energy are shown in Fig.~\ref{fig:selfl}, with the expressions:
\begin{align}
\Sigma^{\ref{fig:selfl}(a)}(\kk,\wt) &= A_{1\kk z} (\ut_{\kk z}^{2} + \vt_{\kk z}^{2}) + 2B_{1\kk z} \ut_{\kk z} \vt_{\kk z} +C_{\kk z}, \\
\Sigma^{\ref{fig:selfl}(b)}(\kk,\wt) &= \frac{4\Jb^{2}}{\wt-2\Ap} \left[ R_{2z} \gk^{2} (1-\gk) (\ut_{\kk z}+\vt_{\kk z})^2  \right.
+\frac{\ut_{\kk z}^2}{2d} (1-\gk) + 2R_{3z}\gk (\ut_{\kk z}^{2} + \ut_{\kk z}\vt_{\kk z}) \nonumber \\
&\left.~~~~~~~~~~~~~-2\gk R'_{3z}(\kk) (\ut_{\kk z}^{2} + \ut_{\kk z}\vt_{\kk z}) \right],  \\
\Sigma^{\ref{fig:selfl}(c)}(\kk,\wt) &= -\frac{4\Jb^{2}}{\wt+2\Ap} \left[ R_{2z} \gk^{2} (1-\gk) (\ut_{\kk z}+\vt_{\kk z})^2  \right.
+\frac{\vt_{\kk z}^2}{2d} (1-\gk) + 2R_{3z}\gk (\vt_{\kk z}^{2} + \ut_{\kk z}\vt_{\kk z}) \nonumber \\
&\left.~~~~~~~~~~~~~-2\gk R'_{3z}(\kk) (\vt_{\kk z}^{2} + \ut_{\kk z}\vt_{\kk z}) \right],  \\
\Sigma^{\ref{fig:selfl}(d)}(\kk,\wt) &= \frac{2\Jb^{2} \gk^{2}}{\wt-2\A} (1-\gk) (\ut_{\kk z}+\vt_{\kk z})^2 R_{2a}, \\
\Sigma^{\ref{fig:selfl}(e)}(\kk,\wt) &= -\frac{2\Jb^{2} \gk^{2}}{\wt+2\A} (1-\gk) (\ut_{\kk z}+\vt_{\kk z})^2 R_{2a}, \\
\Sigma^{\ref{fig:selfl}(f)}(\kk,\wt) &= \Sigma^{\ref{fig:selfl}(g)}(\kk,\wt) =
-\frac{\qq^{2} J^{2}}{\Ap} \Lm_{0}^4 \gk R_{3z} (\ut_{\kk z}+\vt_{\kk z})^2, \\
\Sigma^{\ref{fig:selfl}(h)}(\kk,\wt) &= \frac{2 \qq^{2} J^{2}}{\wt-\Ap-2\A} \left[ \frac{\ut_{\kk z}^{2}}{2d}
+2\gk R_{3a} \Lm_{0}^2 (\ut_{\kk z}^{2} + \ut_{\kk z}\vt_{\kk z})
+\gk^2 R_{2a} \Lm_{0}^4 (\ut_{\kk z}+\vt_{\kk z})^2 \right], \\
\Sigma^{\ref{fig:selfl}(i)}(\kk,\wt) &= -\frac{2 \qq^{2} J^{2}}{\wt+\Ap+2\A} \left[ \frac{\vt_{\kk z}^{2}}{2d}
+2\gk R_{3a} \Lm_{0}^2 (\vt_{\kk z}^{2} + \ut_{\kk z}\vt_{\kk z})
+\gk^2 R_{2a} \Lm_{0}^4 (\ut_{\kk z}+\vt_{\kk z})^2 \right], \\
\Sigma^{\ref{fig:selfl}(j)}(\kk,\wt) &= \frac{2 \qq^{2} J^{2}}{\wt-3\Ap} \Lm_{0}^4
\left[ \gk^2 R_{2z} (\ut_{\kk z}+\vt_{\kk z})^2 + \frac{\ut_{\kk z}^{2}}{2d}
+2\gk R_{3z} (\ut_{\kk z}^{2} + \ut_{\kk z}\vt_{\kk z})\right], \\
\Sigma^{\ref{fig:selfl}(k)}(\kk,\wt) &= -\frac{2 \qq^{2} J^{2}}{\wt+3\Ap} \Lm_{0}^4
\left[ \gk^2 R_{2z} (\ut_{\kk z}+\vt_{\kk z})^2 + \frac{\vt_{\kk z}^{2}}{2d}
+2\gk R_{3z} (\vt_{\kk z}^{2} + \ut_{\kk z}\vt_{\kk z})\right].
\end{align}

\end{widetext}



\begin{thebibliography}{99}
\bibitem{corrfoot}
As explained in Ref.~\onlinecite{i}, the expectation values of the local pair-creation operators $\sum_{i} \langle \tilde{t}_{i\alpha}^{\dagger} \tilde{t}_{i\alpha}^{\dagger} \rangle/N$ are finite and of order $1/d$ within the harmonic approximation, but receive a perturbative $1/d$ correction such that $\langle \tilde{t}_{i\alpha}^{\dagger} \tilde{t}_{i\alpha}^{\dagger} \rangle=0$ to order $1/d$, as required by the constraint. On general grounds, we expect this vanishing to occur order by order in the $1/d$ expansion.
\end{thebibliography}
\end{document}